\documentstyle[eqsecnum,aps,preprint]{revtex}
\begin{document}
\draft
\title{
Structure and shear response in nanometer thick films}
\author{Peter A. Thompson}
\address{
Department of Mechanical Engineering and Materials Science,
and Center for Nonlinear Dynamics and Complex Systems,
Duke University, Durham, NC  27708-0300
}
\author{Mark O. Robbins}
\address{
Department of Physics and Astronomy, The Johns Hopkins University,
Baltimore, MD 21218
}
\author{Gary S. Grest}
\address{
Corporate Research Science Laboratories,
Exxon Research and Engineering Co., Annandale, NJ 08801
}
\maketitle
\begin{abstract}
Simulations of the structure and dynamics of fluid films confined to a
thickness of a few molecular diameters are described.
Confining walls introduce layering and in-plane order in the adjacent fluid.
The latter is essential to transfer of shear stress.
As the film thickness is decreased, by increasing pressure or decreasing the
number of molecular layers, the entire film may undergo a phase transition.
Spherical molecules tend to crystallize, while short chain molecules
enter a glassy state with strong local orientational and translational order.
These phase transitions lead to dramatic changes in the response of
the film to imposed shear velocities $v$.
Spherical molecules show an abrupt transition from Newtonian response
to a yield stress as they crystallize.
Chain molecules exhibit a continuously growing regime of non-Newtonian behavior
where the shear viscosity drops as $v^{-2/3}$ at constant normal load.
The same power law is found for a wide range of parameters, and extends
to lower and lower velocities as a glass transition is approached.
Once in the glassy state, chain molecules exhibit a finite yield stress.
Shear may occur either within the film or at the film/wall interface.
Interfacial shear dominates when films become glassy and
when the film viscosity is increased by increasing the chain length.
\end{abstract}
\pacs{PACS numbers: 68.15.+e, 81.40.Pq, 64.70.Pf, 62.10.+s }
% Liquid thin films; Friction,lubrication, and wear; Glass transition;
% Mechanical properties of liquids.

\section{Introduction}
      \label{sec:intro}

The rheological behavior of fluids in highly confined
geometries is a subject of immense technological importance.
Knowledge of the behavior of confined fluids
is crucial to understanding flows and phase transitions in porous media,
the stability and dynamics of coatings, and
friction and wear in boundary lubrication and nanomachines.
Our knowledge of such thin films has increased greatly in
recent years due to the development of faster computers and
new experimental techniques.
Examples of the latter include atomic force microscopy \cite{AFM1,AFM2},
the quartz crystal microbalance \cite{Krim90},
and the surface forces apparatus (SFA)
\cite{Jacob88a,McGuiggan89,Gee90,VanAlsten88,VanAlsten90a,Hu91,Granick92}.

In this paper we use molecular dynamics simulations to study the structure
and dynamics of fluids in a geometry that models the SFA
\cite{Jacob88a,McGuiggan89,Gee90,VanAlsten88,VanAlsten90a,Hu91,Granick92}.
In the SFA, fluids are confined between mica plates that are atomically
flat over contact diameters of order 100$\mu$m.
The separation $h$ between the plates can be varied with angstr\"om
resolution from contact to many nanometers,
and is monitored using optical interferometry.
Normal and shear forces on the film are measured as the plate
separation is decreased, or as the plates are sheared past each other
at velocity $v$.

SFA experiments have focused primarily on films of
simple hydrocarbon liquids, such as cyclohexane, dodecane,
tetradecane, hexadecane, and
the silicone-liquid octamethlycyclotetrasiloxane (OMCTS)
%% FOLLOWING LINE CANNOT BE BROKEN BEFORE 80 CHAR
\cite{Jacob88a,McGuiggan89,Gee90,VanAlsten88,VanAlsten90a,Hu91,Granick92,Jacob90,Carson92}.
Cyclohexane and OMCTS are roughly spherical molecules,
while dodecane, tetradecane, and hexadecane are short, symmetric chains.
Other fluids that have been studied include the branched molecules squalane and
isoparaffin 2-methyloctadecane, and polymers,
such as polydimethylsiloxane (PDMS) and perfluoropolyether (PFPE)
\cite{VanAlsten90b,disks}.
The latter is commonly used in the computer industry as a lubricant
for thin film magnetic disks where operating conditions typically
require film thicknesses of order $1-5$nm.

These experiments have shown that the static and dynamic properties of
fluids in highly confined geometries can be remarkably different from bulk
properties.
In most cases, bulk behavior persists until separations of order
ten molecular diameters.
At smaller separations there are pronounced oscillations in the
normal load as a function of film thickness \cite{Horn81,Jacob91}.
These are evidence of strong layering within the
film\cite{Abraham78,Toxvaerd81}.
Changes in the dynamic response are even more dramatic.
The effective viscosities $\mu$ of thin films rise more than
five orders of magnitude above bulk values.
Relaxation times are fractions of a second rather than nanoseconds.
When films are sheared rapidly, a wide variety of molecules exhibit
the same power law decrease of viscosity with velocity,
$\mu \propto v^{-2/3}$ \cite{Hu91,Granick92,Carson92}.
In some cases, the response of thin films becomes solid-like:
shear stresses do not relax to zero and there is a substantial
yield stress \cite{Jacob88a,Gee90,VanAlsten88,VanAlsten90a}.
Even those long-chain and branched molecules which do not exhibit layering,
still develop a yield stress after long equilibration or shearing.
When the yield stress is exceeded, ``solid'' films often exhibit
oscillatory stick-slip dynamics\cite{Jacob88a,Gee90,VanAlsten90a}.

Computer simulations have proved to be an effective tool in
understanding the origin of some of these experimental observations.
In addition to revealing the layering that was inferred from
experiment and analytic work
\cite{Horn81,Abraham78,Toxvaerd81,Plischke86,Bitsanis90b,Bitsanis90a},
simulations show that
the periodic potential of the wall lattice induces in-plane order in
the adjacent fluid
\cite{Heinbuch89,Thompson90b,Schoen88,Schoen89,Landman89}.
The degree of in-plane order was found to correlate with the
viscosity at the solid/fluid interface and thus to determine the boundary
condition for fluid flow \cite{Thompson90b,Cieplak94}.
As the film thickness decreases,
confinement can induce a phase transition to a crystalline or glassy phase
%% FOLLOWING LINE CANNOT BE BROKEN BEFORE 80 CHAR
\cite{Schoen89,Thompson90a,Thompson92,Thompson93a,Thompson93b,Robbins93,vanSwol91,Bitsanis93}.
Near the glass transition, films exhibit a
non-Newtonian response with the same power law viscosity seen in experiments
\cite{Hu91,Granick92,Thompson92,Thompson93a,Thompson93b,Robbins93,Manias95}.
Studies of crystalline films of spherical molecules showed that stick-slip
motion resulted from periodic phase transitions between sliding fluid and
static solid phases \cite{Thompson90a,Thompson93b}.
This suggested that the pervasive phenomenon of stick-slip motion is
generally associated with phase transitions between distinct sliding
and static states.

In this paper we take a deeper look at the effect of molecular geometry on
the structure and dynamics of thin films.
We present results for both spherical molecules and freely jointed,
linear chain molecules of varying length $n$.
We first describe the changes in equilibrium structure and diffusion
that occur as films enter a glassy or crystalline state.
The transition pressure and temperature are shifted from bulk values
by confinement.
In some cases the nature of the transition is also different than in the bulk,
i.e. the system enters a glassy rather than crystalline state.
We next consider the dynamic response to shear.
Changing the wall/fluid interaction parameters leads to very different
types of flow profile: shear at the wall/fluid interface, uniform
shear within the film, and shear between layers that are strongly
adsorbed onto each wall.
However,
in each case we find the same power law drop in viscosity with velocity
at constant normal load, $\mu \propto v^{-2/3}$.
Simulations at constant film thickness produce a less rapid fall in viscosity.
Data for the diffusion and relaxation times are fit to the free volume
theory for glass transitions \cite{Grest81}.
Results for slower velocities and higher pressures than studied previously
\cite{Thompson92,Thompson93a,Thompson93b,Robbins93},
reveal that the glassy phase exhibits a yield stress
(implying $\mu \propto v^{-1}$).
This has also been observed in recent experiments \cite{Reiter}.

The organization of the rest of the paper is as follows.
The next section describes the simulation techniques
used in our study.
We then present results that illustrate
the effect of confinement on
the equilibrium properties of spherical and chain molecules.
Section IV examines the corresponding changes in the dynamic
response of films to an imposed shear velocity.
Section V contains a brief summary and discussion.

\section{Simulation Geometry and Method}

In order to mimic the SFA,
our simulations were performed in a planar, Couette geometry
(Fig.\ \ref{fig:snapshot}).
A thin film of spherical or short-chain molecules was confined
between two solid plates parallel to the $xy$ plane.
Edge effects were minimized by applying
periodic boundary conditions within this plane.
Each wall consisted of $2N_w$ atoms forming two [111] planes of an fcc
lattice with nearest neighbor spacing $d$.
Since mica is much more rigid than the confined films, we simplified
the simulations described below by fixing wall atoms to
their lattice sites.
Previous work \cite{Thompson90b}
shows that allowing the wall atoms to
move does not produce qualitative changes in the behavior of the film.
The main effect is to produce a small decrease in the ability of the wall
to order the adjacent fluid at a given temperature $T$ and pressure $P$,
and a corresponding decrease in interfacial viscosity.
The definition of the film thickness $h$ becomes ambiguous at the atomic level.
As shown in the figure and discussed below,
$h$ was defined to exclude the volume occupied by wall atoms.

The chain molecules comprising the fluid were simulated using a
well-studied bead-spring model \cite{Thompson92,Kremer90}.
Monomers separated by distance $r$ interacted through
a truncated Lennard-Jones (LJ) potential,
\begin{equation}
V^{LJ}(r) = \left\{ \begin{array}{ll}
                 4\epsilon [(\sigma/r)^{12} - (\sigma/r)^{6} ] \, , &
                                   r < r_c \\
                 0 \, & r > r_c \, ,
                 \end{array}
         \right.
\end{equation}
characterized by energy and length scales $\epsilon$ and $\sigma$,
respectively.
In some simulations the interaction was made purely repulsive
by setting the cutoff radius $r_c /\sigma =2^{1/6}$.
This minimized the computation time and the temperature dependence.
In other simulations $r_c /\sigma = 2.2$.
Increasing $r_c$ decreases the pressure needed to reach a given density,
and increases the viscosity at that density.
These changes do not alter the nature of the transitions in structure and
dynamics that we describe below.
However, they do shift the location of transitions in the ($P, T$) phase
diagram.

The number $n$ of monomers in a chain was varied from 1 (spherical molecules)
to 24.
Adjacent monomers along each chain were coupled
through an additional strongly attractive, anharmonic potential,
\begin{equation}
V^{CH}(r) = \left\{ \begin{array}{ll}
                 -{1 \over 2} kR_o^2 \ln [1-(r/R_o)^2] \, , & r<R_o \\
                 \infty \, , & r \geq R_o \, .
                 \end{array}
         \right.
\end{equation}
with $R_o=1.5\sigma$ and $k=30\epsilon \sigma^{-2}$.
These values of $R_o$ and $k$ have been used in previous studies
\cite{Thompson92,Kremer90}
of polymer melts at comparable density and temperature.
They were shown to eliminate unphysical bond crossings and bond-breaking.
Another important aspect of this choice of parameters was that $k$ was
not so strong that there was a separation of time scales between
motions induced by $V^{LJ}$ and $V^{CH}$.

Wall and fluid atoms also interacted with a LJ potential,
but with different energy and length scales: $\epsilon_w$,
$\sigma_w$ and $r_{cw}$.
Increasing the ratio $\epsilon_w / \epsilon$, increases the
viscosity of the wall/fluid interface relative to that within the fluid.
The effective viscosity at the interface is also strongly affected
by the relative size of wall and fluid atoms \cite{Thompson90b}.
When the sizes are equal, it is easy for fluid atoms to lock
into epitaxial registry with the substrate, and the viscous coupling
is maximized.
When the sizes are mismatched, the coupling is weaker.
Chain molecules have two characteristic sizes.
The intermolecular potential $V^{LJ}$ has a minimum at $2^{1/6}\sigma$,
while $V^{CH}$ reduces the intramolecular separation
to about 0.96$\sigma$ in our simulations.
The presence of an extra length scale frustrates epitaxial
order as discussed below.

The simulations were performed in an ensemble where the number of
monomers $N_f$, the system temperature $T=1.1 \epsilon/k_B$,
and the normal pressure
or load $P_{\bot}$ exerted on the top wall were all constant.
The latter constraint allowed the plate separation
$h$ to vary, as in experiments.
To maintain constant load in the $z$ direction, we added
the following equation of motion for the top wall:
\begin{equation}
\ddot{Z} = (P_{zz}-P_{\bot})A/M \, ,
\label{eq:Z}
\end{equation}
where $Z$ is the average $z$ coordinate of atoms in the top wall,
$P_{zz}$ is the $zz$-component of the instantaneous, microscopic
pressure-stress tensor \cite{Allen87} at the wall, and
$A$ is the area of the wall.
The mass $M$ is an adjustable parameter that controls the oscillation
frequency of the wall.
A small $M$ leads to unphysically rapid oscillations,
while a large $M$ leads to a slow exploration
of the parameter space associated with $h$.
The times associated with the actual mass of mica sheets in an SFA
are prohibitively long.
We found that our results were relatively insensitive to $M$ as
long as the period of height oscillations was longer than that
of the longest phonon period in our finite fluid films.
This condition was satisfied in the simulations described below
by setting $M$ equal to twice the product
of the monomer mass and the number of wall atoms.

Shear was imposed by pulling the top plate at a fixed velocity $v$.
The shear stress was obtained from two independent and consistent
methods: direct calculation of the forces exerted on the walls by
fluid molecules,
or from the $xz$ component of the the microscopic pressure-stress tensor
\cite{Allen87}.
In the SFA, the transverse alignment of walls
in the y direction is not rigidly fixed.
The top wall can displace in this direction as it slides
in response to stress from the film.
We incorporated this motion in most of the simulations described below
by coupling the $y$ degree of freedom of the top plate
to a spring of force constant $\kappa_y$.
The extra equation of motion is:
\begin{equation}
\ddot{Y} = (P_{yz}A -\kappa_y Y) /M \, ,
\label{eq:Y}
\end{equation}
where $Y$ is the average $y$ coordinate of atoms in the top wall.
The transverse stress, $P_{yz}$ can generally be minimized by displacements
of order $d$.
We used a value of $\kappa_y/M = 0.05\epsilon m^{-1} \sigma^{-2}$
that was small enough to allow such
displacements, and large enough to prohibit displacements that were
comparable to the system size.
Simulations with variable $Y$ gave slightly smaller shear stresses
than those with fixed $Y$,
but did not change the scaling of shear stress with shear rate.

Constant temperature was maintained by coupling the $y$ component
of the velocity to a thermal reservoir \cite{Grest86}.
Langevin noise and frictional terms were added to the equation of motion
in the $y$ direction,
\begin{equation}
m\ddot{y} = -{ \partial \over {\partial y}} (V^{LJ}+V^{CH})
            -m\Gamma\dot{y} + W \, ,
\label{eq:therm}
\end{equation}
where $\Gamma$ is the friction constant that controls the rate of heat
exchange with the reservoir, and
$W$ is the Gaussian distributed random force from the heat bath
acting on each monomer. Note that the variance of $W$ is related
to $\Gamma$ via the fluctuation-dissipation theorem,
\begin{equation}
<W(t)W(t^\prime)>=2mk_B T\Gamma\delta(t-t^\prime) \, .
\end{equation}
If $\Gamma$ is too small, the energy dissipated in shearing the system
will cause the temperature to rise.
If $\Gamma$ is too large, the random force is not integrated accurately.
We found the system was well-behaved with $\Gamma=2\tau^{-1}$,
where $\tau = (m \sigma^2/\epsilon)^{1/2}$ is the characteristic LJ time.
The equations of motion were integrated using a fifth-order Gear
predictor-corrector algorithm with a time step $\Delta t = .005 \tau$.

The major difference between our theoretical ensemble and experiment
is the constraint of constant $N_f$.
In experiments, fluid can drain from the compressed region between
the mica plates to equilibrate with the external chemical potential.
However, the equilibration time may be extremely long due to the
small ratio between $h$ ($\sim 1$nm) and the contact diameter
($\sim 100 \mu$m) and the immense viscosities of confined films
\cite{Horn89}.
Equilibration is more likely in steady sliding experiments
where the contact area changes completely \cite{Jacob88a,Gee90,Jacob90},
than in the experiments with small ($\sim 10$nm)
sinusoidal motions that we compare to here
\cite{VanAlsten88,Granick92,Reiter}.

To determine $N_f$ we used the fact that the film orders into
well defined layers of fairly constant density $\rho_L$.
We set $N_f = m_l A \rho_L$ where $m_l$ was the desired number
of layers.
This guaranteed that the nominal density of films was independent
of the number of layers.
Equilibrium configurations were obtained through a sequence of runs.
The load was decreased to a small value for $10^4 \Delta t$,
so that chains could move rapidly.
The load was then increased sequentially, allowing at least
$10^5 \Delta t$ for equilibration.
To check that equilibrium was reached, we compared runs with
different equilibration times and starting configurations.
We also compared results with increasing and decreasing load,
and runs equilibrated at fixed load and high temperature.
The film thickness $h$ was defined in terms of the distance $h_0$ between the
innermost wall layers as $h=h_0 m_l/(m_l+1)$.
This is equivalent to subtracting a distance of half the layer spacing
from $h_0$ for each wall to account for the volume taken up by wall atoms.
Other methods of correcting for the size of wall atoms give essentially
the same results.
At most the viscosity is changed by a constant multiplicative factor of
order unity.

\section{Equilibrium Structure }

Two types of structure are induced in the fluid adjacent
to a flat solid surface:
Layering normal to the surface and epitaxial order in the plane
of the surface.
When a fluid film is confined between two solid surfaces,
both types of order may span the entire film.
Experimental consequences are oscillations in the normal force
with film thickness \cite{Horn81}, and phase transitions
to solid states that resist shear forces
\cite{Jacob88a,Gee90,VanAlsten88,VanAlsten90a}.

Layering has been studied extensively for a variety of
molecular geometries
%% FOLLOWING LINE CANNOT BE BROKEN BEFORE 80 CHAR
\cite{Abraham78,Toxvaerd81,Plischke86,Bitsanis90b,Bitsanis90a,Heinbuch89,Thompson90b,Schoen88,Bitsanis93,Magda85,Kumar88,Ribarsky92,Xia92}.
It is induced by the monomer pair correlation function $g(r)$,
and the sharp cutoff in fluid density at the wall.
Fig. \ref{fig:density} shows plots of the density as a function of
the distance between walls for monomers and 6-mers.
Note the well-defined density peaks near the walls, and the
decay of oscillations into the center of the film.

In general,
the sharpness and height of the first density peak are determined
mainly by the wall/fluid interaction, and increase with
$\epsilon_w$ or $P_{\bot}$.
The rate at which the density oscillations decay is determined by the
decay of correlations in g(r).
Except near critical points, the decay is a few molecular diameters.
For fluids of simple spherical molecules, pronounced density oscillations
can extend up to $\sim 5\sigma$ from an isolated solid surface.
The competition between intra- and inter- molecular spacings leads
to less pronounced layering with chain molecules (Fig. \ref{fig:density}).
The degree of layering is fairly
independent of chain length once $n$ exceeds 6.
Simulations with realistic potentials for alkanes show more pronounced
layering near the wall due to a transition to an extended state of the
chains in the first layer \cite{Ribarsky92,Xia92}.

In-plane order is induced by the in-plane variation, or corrugation,
in the potential from wall atoms.
As for layering, the rate of decay is determined by g(r).
It has been studied extensively for spherical molecules
\cite{Abraham78,Toxvaerd81,Heinbuch89,Thompson90b,Schoen89,Landman89},
but only recently for chain molecules \cite{Manias95,Manias93,Manias94}.
In part, this is because most studies of chain molecules near surfaces
have neglected the discrete lattice structure of the solid surface.
In-plane order plays an essential role in determining the dynamics
of fluids near solid interfaces.
For example, we have shown that flow boundary conditions
near solids are well correlated with the amount of in-plane order,
and not with the degree of layering \cite{Thompson90b,Cieplak94}.

The amount of in-plane order depends upon the
relative size of wall and fluid atoms, as well as the strength
of their interaction \cite{Thompson90b}.
If the spacing between fluid atoms is equal to that of wall
atoms, the atoms can lock in to all the local minima in
the wall potential.
In an earlier study of simple spherical molecules \cite{Thompson90b}
we found crystallization of one or two fluid layers adjacent to
a solid interface.
If the wall and fluid atoms have different sizes,
epitaxial order is frustrated \cite{Thompson90b}.
As we now show, the presence of two different length scales in our
chain molecules also frustrates in-plane order.

Two measures were used to quantify the degree of in-plane ordering.
The first was the spatial probability distribution
$\rho_l(x,y)$ of monomers in the $l$th layer relative to the unit cell of
the solid lattice.
The second was the two-dimensional static structure factor
$S(k_x, k_y)$ evaluated in the $l$th layer according to
\begin{equation}
S_l (\vec{k}) = 1/N_l { \left| \sum_{i=1}^{N_l} e^{i\vec{k}\cdot\vec{r}_i}
                        \right| }^2 \, ,
\label{eq:struct}
\end{equation}
where $N_l$ was the number of monomers $i$ within the layer.

Figures \ref{fig:epi1} and \ref{fig:epi6} show $\rho(\vec{r})$ and
$S(\vec{k})$ for a two layer film at two loads.
Values of $S$ and $\rho$ for each layer were averaged for $250\tau$.
Then results for layers related by symmetry about the middle
of the film were averaged to improve statistics.
At the lowest applied load, $P_{\bot}=4 \epsilon \sigma^{-3}$,
there is little evidence of any in-plane order in $\rho$ for either $n=1$ or 6.
The structure factor $S$ shows weak rings that are characteristic of liquid
order, and small peaks at the reciprocal lattice vectors
$\vec G_i$ of the solid surface.
These peaks represent a linear response to the surface potential.
Their strength is best expressed \cite{Thompson90b}
in terms of the Debye-Waller factor $e^{-2W} \equiv S(G_0)/N_l$
at the smallest reciprocal lattice vector $\vec G_0$.
The Debye-Waller factor is unity in an ideal crystal at zero temperature.
Thermal fluctuations decrease $e^{-2W}$ to about 0.6 at the melting point
of a bulk crystal \cite{Stevens93}.
The value of $e^{-2W}$ at $P_{\bot}=4 \epsilon \sigma^{-3}$
is only 0.06 for both $n=1$ and 6.
Such small values are clear evidence of liquid structure.

As the load increases, the walls induce more in-plane order.
At $P_{\bot}=16\epsilon \sigma^{-3}$,
there are well-defined modulations in $\rho(\vec{r})$.
The peaks in $\rho$ are located above gaps in the adjacent solid layer.
For $n=1$, the monomers are nearly always over these gaps.
Less order is induced in films of chain molecules because of the
different intra- and inter-molecular spacings.
Note the pronounced ridges that connect the peaks in $\rho(\vec{r})$.
These follow the lowest energy path between the gaps in wall atoms.

The monomer results for
$S(\vec{k})$ and $\rho(\vec{r})$ clearly indicate crystalline order at
$P_{\bot}=16 \epsilon \sigma^{-3}$.
For example, the mean-squared displacement of a fluid atom from
the peaks in $\rho(\vec{r})$ is $\sim 0.017 d^2$, which is below the
Lindemann criterion for melting \cite{Stevens93,Lindemann}.
Furthermore, $S(\vec{k})$ shows many sharp Bragg peaks,
including several higher-order harmonics.
The Debye-Waller factor, $e^{-2W}=0.78$, is above the bulk
melting criterion \cite{Stevens93} and is consistent with the mean-squared
displacement.
In contrast, the chain molecules never crystallize within our
simulation times.
The degree of order increases with $P_{\bot}$ and then saturates.
The Debye-Waller factor is only $0.43$ in Fig. \ref{fig:epi6}.

Figure \ref{fig:loaddep} shows the load dependence of film thickness,
Debye-Waller factor and in-plane diffusion constant $D$ in 2 layer
films \cite{Thompson92}.
Note the sharp transitions in the monomer results at
$P_{\bot}=7 \epsilon \sigma^{-3}$.
At lower loads the structure is liquid-like and diffusion is rapid.
At higher loads the film has crystallized and diffusion is suppressed.
The lowest value of $e^{-2W}$ within the crystalline phase (0.62) is
very close to that in bulk crystals at their melting point \cite{Stevens93}.
There is a similar phase transition in bulk films, but it occurs at
$P_{\bot}=12 \epsilon \sigma^{-3}$.
Using typical values for $\epsilon = 200K$ and $\sigma = 3$\AA,
this corresponds to a shift of 0.5 GPa.
Note that the absolute loads are sensitive to the potential cutoffs
$r_c$ and $r_{cw}$,
but the shift in the transition point is less so.

Results for chain molecules show more gradual changes \cite{Thompson92}.
The Debye-Waller factor saturates at about 0.4, and remains nearly
unchanged if the load is increased by an additional order of magnitude.
The only evidence of a transition to a new phase comes from measurements
of dynamic quantities, such as the diffusion constant.
The value of $D$ drops rapidly below that in a bulk
fluid at the same pressure.
Studies of the viscous response, described in the next section,
also show a dramatic slowing of molecular motions.
These findings are evidence that the film undergoes a glass
transition.
Extrapolation to $D =0$ using free volume theory indicates a
transition near $P_{\bot}^G=16\epsilon \sigma^{-3}$ (Fig. \ref{fig:gammac}).
At loads above this value we find that films exhibit solid behavior
over the time scales of our simulations:
there is no relaxation of applied shear forces and diffusion is
undetectably small.
Note that diffusion in bulk films extrapolates to zero at
a much higher pressure, $P^G_{bulk} \approx 24 \epsilon \sigma^{-3}$.
As for spherical molecules, confinement produces large shifts ($\sim 0.8GPa$)
in the bulk transition pressure.
These shifts are comparable to the glass transition pressures
($\sim$0.5GPa) of bulk lubricants at room temperature \cite{Alsaad78},
and one may expect that many lubricants vitrify in thin films.

Although chain molecules do not crystallize, there are pronounced
changes in their configuration and orientation as they approach the
glass transition.
Figure \ref{fig:length} shows the distribution of end-to-end distances
$R_{1n}$ for $n=6$ at three different pressures.
At the lowest load, $P_{\bot}=1 \epsilon  \sigma^{-3}$,
the film thickness $h= 2.94 \sigma$ is larger than the radius of
gyration of a free three-dimensional chain \cite{foot3d}, and
the density is nearly independent of $z$.
As a result the distribution of end-to-end distances is close to
that for an ideal Gaussian chain \cite{Flory69}.
The major difference is a cutoff in the distribution at small
separations due to the hard core repulsion between monomers.
As $P_{\bot}$ increases, the distribution becomes more structured,
indicating that monomers are being locked into a discrete set
of locations.
One also sees that highly stretched configurations become less likely.
Previous studies of polymers confined to two dimensions show
that they maximize their entropy when each polymer segregates
into its own region of the plane \cite{Carmesin90}.
The result is a collapse into a compact configuration of the chain.
As $P_{\bot}$ increases from 1 to 16$\epsilon \sigma^{-3}$, $h$ drops from
a value greater than the radius of gyration to a smaller
value, and the behavior crosses over from three- to two-dimensional
\cite{foot3d}.

Figure \ref{fig:theta} shows how the orientation
of intramolecular bonds varies with load.
We use the conventional coordinate system where $\theta$ is the
polar angle relative to the $z$ axis and $\phi$ is the azimuthal
angle within the $xy$ plane and relative to the $x$ axis.
Since the sequence along the chain is arbitrary, the sign of the
vector connecting nearest neighbors is not defined.
Thus we calculated the distribution of $\cos^2 \theta$
for all intra-molecular bonds.
The distribution for completely random orientations,
$1/(2\cos \theta)$, is shown by dashed lines in the plots.
At $P_{\bot}=4\epsilon \sigma^{-3}$, the distribution is nearly random.
For $P_{\bot}=16 \epsilon \sigma^{-3}$ one finds sharp peaks at preferred
orientations.
The suppression of other orientations limits the ability of molecules
to realign and results in a decreased diffusion rate.
At the preferred values of $\cos^2 \theta=0$ and 0.6,
bonds connect monomers in the same ($\theta=90^\circ$)
or adjacent ($\theta \approx 40^\circ$) layers.
Similar peaks appear in the distribution of $\cos^2 \phi$
near 0 and 0.75,
or $\phi = 90$ and 30$^\circ$.
These are the directions of bonds between monomers that have locked
epitaxially into minima in the wall potential or lie along the density
ridges in Fig. \ref{fig:epi6}.

The structural changes in 2 layer films of chains with $n > 6$ are nearly
the same as the 6-mer results just described.
In general, the structure becomes insensitive to chain length when
the film thickness is much smaller than the radius of
gyration in a bulk melt.
As we now show, the shear response of the system may also become
insensitive to $n$ in this limit.

\section{Response to steady shear }

SFA experiments probe the dynamic response of films by measuring
the shear force on the walls as a function of velocity
\cite{Gee90,Granick92,Georges}.
This measurement necessarily combines information about the
shear response of the film and the interface.
There is no way to determine whether shear occurs primarily
within the film or is localized at the interface.
It has also been impossible to examine structural changes
in shearing films.

Experimentalists generally assume a no-slip boundary condition
at the wall/fluid interface, i.e. that all shear occurs within the film.
The shear rate $\dot\gamma \equiv \partial v_x/\partial z$ is
then given by $v/h$.
An effective viscosity of the film $\mu$ can be determined
using the macroscopic relation between force and shear rate in
a lubricant film
\begin{equation}
f= \mu \dot \gamma = \mu v/h
\label{eq:mu}
\end{equation}
where $f$ is the force per unit area or shear stress.
In a simple Newtonian fluid, $\mu$ is independent
of shear rate and $f$ rises linearly with $v$.
In most non-Newtonian fluids, $\mu$ decreases with $v$,
and $f$ rises sublinearly \cite{Graessley74,Ferry80}.

Typical force-velocity curves for chain molecules are shown in
Fig. \ref{fig:fvsv}, where $n=6$, $\epsilon_w = \epsilon$ and $m_l=2$.
The top wall was sheared at constant velocity $v$ in the $x$ direction,
and allowed to adjust its registry in the $y$ direction in response
to shear stresses as described above.
Results for $P_{\bot} = 8$ and $12 \epsilon \sigma^{-3}$ rise rapidly
from zero and then saturate as $v$ increases.
The film is highly non-Newtonian and, as shown below,
$f$ scales roughly as $v^{1/3}$.
At $P_{\bot} = 16$ and $20 \epsilon \sigma^{-3}$,
the force approaches a constant value or yield stress as $v \rightarrow 0$.
This behavior is characteristic of solid-on-solid
friction \cite{Bowden58}, and is further evidence that the film is in a
glassy state at these loads.

Granick and coworkers \cite{Hu91,Granick92} have plotted their shear response
curves in terms of the effective viscosity and shear rate defined
in Eq. \ref{eq:mu}.
Fig. \ref{fig:muvsg}(a) shows
the data of Fig. \ref{fig:fvsv} replotted in this manner.
A Newtonian regime with constant viscosity $\mu_0$ is seen
at the lowest loads and shear rates.
As $\dot \gamma$ increases, the system can no longer respond rapidly
enough to keep up with the sliding walls.
When $P_{\bot}$ is below the glass transition load $P_{\bot}^G$,
there is well-defined crossover shear rate $\dot \gamma_c$ at which the
viscosity begins to drop.
The shear-thinning appears to follow a universal power law
$\mu \propto v^{-2/3}$ (implying $f \propto v^{1/3}$) that is indicated
by a dashed line on the figure \cite{Hu91,Granick92,Thompson92}.

The inverse of $\dot \gamma_c$ corresponds to the longest structural
relaxation time of the film \cite{Graessley74,Ferry80}.
As $P_{\bot}$ rises toward $P_{\bot}^G$, $\dot \gamma_c$ drops
to zero and the relaxation time diverges.
This slowing of dynamics in the film is directly related to
the decrease in diffusion constant with $P_{\bot}$ seen in
Fig. \ref{fig:loaddep}, and the rapid rise in
$\mu_0$ with $P_{\bot}$ seen in Fig. \ref{fig:muvsg}.

One common description of glass transitions is the free volume model
\cite{Grest81}.
It states that $\dot \gamma_c$ and $D$ should vanish as
$\exp (-h_0/(h-h_c))$ where $h_c$ is the film thickness at the
glass transition \cite{Thompson93b}.
Fig. \ref{fig:gammac} shows fits of both quantities to
this form with $h_c$ corresponding to a glass transition near
$P_{\bot}^G=16\epsilon \sigma^{-3}$.
There is considerable debate about the proper scaling of dynamics
near a glass transition, and even about the existence of a sharp
transition.
The success of the fit in Fig. \ref{fig:gammac} should be taken
as evidence that the transition we see is like other glass transitions,
rather than evidence for the free volume theory.

When $P_{\bot}$ exceeds $P_{\bot}^G$, there is a
qualitative change in the response at small $\dot \gamma$.
As shown in Fig. \ref{fig:muvsg}, $\mu$ falls more steeply
with $v$ at $P_{\bot} = 16$ and $20 \epsilon \sigma^{-3}$.
The slope on this log-log plot approaches $-1$ as $\dot \gamma$
decreases.
This implies a constant shear force or yield stress,
as already seen for these loads in Fig. \ref{fig:fvsv}.
A constant shear force is typical of solid-on-solid sliding
\cite{Bowden58}.
We show below that the film is nearly rigid at these loads,
and that shear is largely confined to the wall/film interface.
This regime was not evident in previous simulations
\cite{Thompson92,Thompson93a,Thompson93b,Robbins93},
because they did not extend to high enough loads and low enough
velocities.

Granick et al. have found a very similar series of changes with load
\cite{Hu91,Granick92,Reiter}.
Indeed they are even more dramatic, because a larger range of time
scales is accessible to experiments.
Increasing the load produced a decrease in $\dot \gamma_c$ by
at least 10 orders of magnitude, and a rise in the Newtonian
viscosity $\mu_0$ by more than 5 orders of magnitude.
At loads where $\dot \gamma_c$ was still non-zero \cite{Hu91},
there was a pronounced non-Newtonian regime where
$\mu$ dropped as $v^{-2/3}$.
At higher loads there is a crossover to constant shear force
\cite{Reiter}.

A decrease in viscosity with increasing shear-rate is normally
accompanied by structural changes that reflect the fact that the system
no longer has time to relax back to the equilibrium state
\cite{Graessley74,Ferry80}.
The main structural change in our constant load simulations
is an increase in the film thickness with shear rate.
Shear produces an extra normal force that separates the walls.
This facilitates sliding and lowers $\mu$.
Note in Fig. \ref{fig:muvsg}(b) that $h$ is constant in the Newtonian
regime for a given load and begins to rise only
when $\dot \gamma$ exceeds the value of $\dot \gamma_c$.
Once the film is in the glassy phase, changes in $h$ extend to the
lowest practical shear rates.
The changes in $h$ may be coupled to structural changes within the film,
but these are difficult to detect until relatively
high shear rates ($\dot \gamma > 0.1 \tau^{-1}$).
In this regime, the number of intramolecular bonds which connect layers
begins to decrease to facilitate flow \cite{Manias95,Manias93,Manias94}.

The above results imply that $\mu$ should drop less rapidly with
$\dot \gamma$ if $h$ is fixed.
This is verified by the constant $h$ data shown in
Figure \ref{fig:fixh}.
We find that $\mu $ decreases with a power law near -0.5 in this ensemble
\cite{Thompson92,Manias95}.
Previous experiments have been done at fixed $P_{\bot}$ and it would be
interesting to test our simulation results against experiments
at fixed $h$.
Studies of structural changes in our fixed $h$ simulations show that
the layers of monomers move away from the walls to facilitate interfacial
shear and the layering at the wall becomes slightly more defined.
As above, there is little structural change within the films.

Figure \ref{fig:Ndep} shows how $f$ depends on chain length in 2 layer
films at $P_{\bot}=16\epsilon \sigma^{-3}$.
Note that the results become independent of chain length for large $n$.
Only the results for $n=1$ and 3 are noticeably different.
This length independence may seem surprising since the bulk viscosity
rises rapidly with $n$.
However, at these high loads, films of long chains have frozen into a glassy
state and all the shear occurs at the interface.
As noted in the previous section,
the structure of the interface and the resulting shear coupling
are insensitive to $n$ when chains are long enough to span the system.

In Fig. \ref{fig:muvsg}, $\mu$ and $\dot \gamma$ were obtained from $f$ and $v$
by assuming a no-slip flow boundary condition (Eq. \ref{eq:mu}).
This assumption is invalid when shear localizes at the interface.
Mechanical equilibrium requires that the shear {\em stress}
be uniform throughout
the system, however the shear {\em rate} may vary with position.
In Fig. \ref{fig:profile} we show the variation of $v_x$ with $z$
for a number of systems.
The actual shear rate within the film,
$\dot \gamma_{film} = \partial v_x / \partial z$, is given by the slope of
the flow profiles.
For the monomer results shown in Fig. \ref{fig:profile}(a), there is
relatively little shear at the walls and
$\dot \gamma_{film} \approx \dot \gamma$.
As $n$ increases to 6 (Fig. \ref{fig:profile} (a)),
shear becomes predominantly confined to the wall/film interface.
Fig. \ref{fig:ratio} shows the variation of
$\dot \gamma _{film} / \dot \gamma$ with $n$ at
$P_{\bot}=16\epsilon \sigma^{-3}$.
This ratio would be unity if a no-slip boundary condition applied.
While the no-slip approximation is very accurate for monomers,
it fails rapidly as $n$ increases to 2 and beyond.
For long chains, only 15\% of the shear occurs within the film at
$P_{\bot}=16 \epsilon \sigma^{-3}$.
This type of flow profile is frequently called plug-like flow.

The power law shear-thinning shown for two layer films in Fig. \ref{fig:muvsg}
is quite universal.
Fig. \ref{fig:muvsg2} shows results for a system where $\epsilon _w$
was increased to $3\epsilon$ in order to decrease the amount of
shear at the interface \cite{Thompson92}.
Note that the glass transition can be approached in two different
ways because the same load can be obtained with different numbers
of layers \cite{Horn81}.
Panel (a) shows results at fixed load with a decreasing number of
layers, and panel (b) shows the effect of
increasing load at $m_l = 4$.
In either case, there is an increase in the Newtonian
viscosity and a decrease in $\dot \gamma_c$ as the fluid
becomes more confined.
The shear-thinning region shows the same power law dependence
seen in Fig. \ref{fig:muvsg} and experiment \cite{Hu91,Granick92}.
If the load is increased further into the glassy regime,
shear will localize at the interface.
In this limit one finds a regime where the shear force is velocity
independent.

A typical flow profile for the system just discussed is shown in
Fig. \ref{fig:profile}(b).
Note that while all the shear occurs within the film, it is
not spread uniformly throughout the film.
Over the entire range of power law shear thinning, shear
is localized at a plane near the center of the film.
Closer examination of particle motions reveals
that each molecule adsorbs tightly to one of the two walls.
These adsorbed films then slide past each other.
Hence, the power law shear thinning is still
associated with interfacial sliding.
The only difference from the system of Fig. \ref{fig:loaddep}
is that the interface is within the film rather than
at the wall film interface.

To determine whether uniform shear could produce the same
power law response, we examined several other sets of
parameters.
Fig. \ref{fig:profile}(b) also shows a flow profile for
$\epsilon_w/\epsilon = 1$, $P_{\bot}=8\epsilon \sigma^{-3}$,
and $r_{cw}=r_c=2.2\sigma$.
Note that the shear rate is constant within the film ($v_x$ is linear).
Roughly half of the shear occurs in the film and the other half at the
two wall/film interfaces.
Fig. \ref{fig:muvsg3} shows the variation of $\mu$ with $\dot \gamma$.
As for other systems, there is a substantial range where
$\mu \propto \dot \gamma ^{-2/3}$.
Thus we conclude that even films with uniform shear exhibit the
same power law near the glass transition.
Note that if the interface and film had different power laws
the flow profile would change with $\dot \gamma$.
We find the same division of shear between interface and film
over the entire range of shear rates.

\section{Discussion and Summary }

The simulations reported here, and other related work,
show that confinement can produce a rich variety
of phenomena in nanometer scale films.
Structural changes, including layering and in-plane order,
are now well-documented
\cite{Toxvaerd81,Bitsanis90b,Bitsanis90a,Thompson90b,Magda85}.
We have shown here, that both types of ordering are smaller
in chain molecules due to the presence of more than one
length scale.
This effect should be even more pronounced in films of branched
or other more complicated molecules.

Confinement may eventually lead to a phase transition within the film
(Fig. \ref{fig:loaddep}) \cite{Schoen89,Thompson92,Thompson93b,Robbins93}.
In some cases, the transition is just a bulk phase transition that
is shifted to a lower pressure or higher temperature.
This is the case for crystallization of spherical molecules by walls whose
atoms have nearly the same size.
In other cases, the nature of the transition is changed.
Chain molecules tend to be trapped in a glassy state.
Spherical molecules can form a glassy phase or a new crystal structure
when the walls are amorphous or epitaxy is frustrated due to size
differences \cite{Thompson90b,Thompson90a}.
Glassy phases also occur when the crystalline axes of the confining walls
are not aligned,
because the walls induce contradictory ordering tendencies on the film
\cite{tobe}.
The mica walls in the SFA are generally not aligned and glassy
phases may be the rule in experimental studies.
This would explain why the same power-law non-Newtonian behavior is
observed in films of chain molecules, like alkanes, and more spherical
molecules, like OMCTS \cite{Hu91,Granick92}.
An interesting observation from our simulations is that
thicker films (3 or 4 layers) are more able to accommodate the conflicting
influences of the walls and may have a higher yield stress
than films with only 1 or 2 layers.
Such films typically melt when they begin to slide and will be the topic
of a later paper.

Three distinct types of dynamic shear response were found.
Newtonian behavior occurs at large thicknesses and temperatures
and at low pressures.
In the opposite limits there is
a constant shear stress or yield stress as $v \rightarrow 0$.
The transition between the two types of behavior occurs abruptly
if the film crystallizes, and gradually if it vitrifies.
The third type of response occurs near the glass transition
where there is an extended range of power law shear thinning
$\mu \propto \dot \gamma ^{-2/3}$ ($f \propto v^{1/3}$).
The power law holds from a crossover shear rate $\dot \gamma_c$
up to the frequencies of typical phonons.
As the glass transition is approached, $\dot \gamma_c$ drops
rapidly to zero.
There is a corresponding drop in the diffusion constant and a rise in the value
of $\mu$ at $\dot \gamma < \dot \gamma_c$.

These types of response describe the behavior of the entire system,
including film and wall/film interfaces.
Although the shear stress is independent of position, the shear rates
within the film and at the interface may be very different.
The shear rate at the interface decreases as the strength of the wall fluid
coupling ($\epsilon_w$ and/or $r_{cw}$) increases,
and when $d$ is such that monomers can easily lock into epitaxy with the wall.
The shear rate within the film depends strongly on $n$, $m_l$ and their
relative sizes.
When chains are strongly adsorbed, shear may become localized at the
midpoint of the film.

All flow profiles seem to be associated with the same
power law shear thinning.
A different power law was obtained only when the ensemble was
changed from constant load to constant thickness.
For constant film thickness $\mu$  decreases as $\dot\gamma^{-1/2}$.
Recent simulation studies of the bulk shear viscosity for the same
molecular potential find a very similar shear thinning exponent
at constant volume \cite{Kroger93}.
It would be interesting to compare these results to bulk simulations
at fixed pressure.

Several models have been developed to explain the power law shear
thinning observed in experiment and simulations
\cite{Rabin93,Urbakh95,degennes95}.
All the models find shear thinning with an exponent of 2/3 in certain
limits.
However, they start from very different sets of assumptions and it
is not clear if any of these correspond to our systems.
Two of the models work at constant film thickness \cite{Rabin93,Urbakh95}
where we find an exponent of 1/2.
One of these \cite{Urbakh95} also finds that the exponent depends on the
velocity profile, while our results do not.
The final model \cite{degennes95} is based on scaling results for
the stretching of polymers under shear.
While it may be appropriate for thick films, it can not describe
the behavior of films which exhibit plug-like flow.
It remains to be seen if the 2/3 exponent has a single explanation
or arises from different mechanisms in different limits.

Shear is known to stretch and align bulk polymers.
One might expect similar changes in thin films, but confinement limits
the ability of polymers to rotate in the $xz$ plane.
It has been suggested that chains might prefer to align normal to
the $xz$ plane in this case \cite{Gee90}.
We found no tendency for shear-induced alignment in any direction
when the film was much thinner than the bulk radius of gyration
and the shear rate was less than $0.1\tau^{-1}$.
Fig. \ref{fig:time} shows the decay of order in a film where all
molecules were originally aligned in the $x$ direction.
The projections of the vector between the ends of the chains on to
the $x$ and $y$ axes are plotted against time.
There is a rapid decrease in anisotropy over $\sim 10 \tau$,
and the ratio of the projections (Fig. \ref{fig:time}(c)) reaches
unity after about $450 \tau$.
There does appear to be memory of the direction of sliding for even longer
times, but this must be reflected in more subtle details
of the structure, such as the relative positions of monomers in the
$xy$ plane.
Studies of the development of orientational order as the degree of
confinement is reduced are underway \cite{Arlet}.

\acknowledgments

We thank S. Granick, J. N. Israelachvili, P. M. McGuiggan and
J. Klein for useful discussions.
Support from National Science Foundation Grant DMR-9110004
and the Pittsburgh Supercomputing Center is gratefully acknowledged.

{\large \bf Figure Captions}
\begin{figure}
\caption{
Projection of particle positions onto the $x-z$ plane
illustrating the simulation geometry.
Periodic boundary conditions are imposed in the $x$ and $y$ directions.
Gray circles connected by lines show the short chain molecules
that make up the film of thickness $h$.
Solid circles indicate wall atoms.
They sit on the sites of fcc [111] layers in the $x-y$ plane.
A constant load $P_\bot$ is applied in the direction normal
to the top wall, and the bottom wall is held fixed.
Shear may be applied by moving the top wall at fixed velocity $v$
in the $x$ direction.
In most simulations, the wall was allowed to optimize its registry
in the $y$ direction as sliding occurred (Eq. \protect\ref{eq:Y})
}
\label {fig:snapshot}
\end{figure}
\begin{figure}
\caption{
Profiles of the monomer number density $\rho$ as a function of $z$ at $v=0$
for films with (a) $n=1$, $h=3.41\sigma$ and (b) $n=6$, $h=3.22\sigma$.
Other parameters were $\epsilon_w/\epsilon=1$, $r_{cw}=r_c=2.2\sigma$, $m_l
=4$,
$P_\bot=8\epsilon \sigma^{-3}$, $k_B T=1.1 \epsilon$, $N_f= 288$, $N_w=72$,
and $d=1.2\sigma$.
The number of monomers in bins of width $h/80$ along the z direction
was averaged over 150$\tau$.
Note that $z$ is normalized by the average plate separation, $h$.
}
\label{fig:density}
\end{figure}
\begin{figure}
\caption{
Values of $\rho(\vec{r})$ (right) and $S(\vec{k})$ (left)
for a two layer monomer film at
$P_{\bot} = 4$ (bottom) and 16$\epsilon \sigma^{-3}$ (top).
The scale for each quantity is adjusted to fill the figure.
All data were averaged over $250 \tau$ with $v=0$, $\epsilon_w/\epsilon =1$,
$r_{cw}=r_c=2^{1/6}\sigma$, $N_w=144$, $N_f=288$, $k_B T=1.1\epsilon$,
and  $d=1.2\sigma$.
Wall atoms lie on the sites of a triangular lattice and are
at the corners and center of the unit cell shown for $\rho$.
Sharp Bragg peaks are found in $S(\vec{k})$ at the reciprocal
lattice vectors of the wall surface.
At $P_{\bot}=4 \epsilon \sigma^{-3}$, the peaks are barely above the
circular ridge that is characteristic of fluid structure.
}
\label {fig:epi1}
\end{figure}
\begin{figure}
\caption{
Results for 6-mers with the same parameters as in \protect\ref{fig:epi1}.
Note the pronounced ridges between peaks in $\rho$ for
$P_{\bot}=16\epsilon \sigma^{-3}$.
}
\label {fig:epi6}
\end{figure}
\begin{figure}
\caption{
Effect of varying load on (a) plate separation,
(b) Debye-Waller factor, and (c) diffusion constant
for motion parallel to the walls.
All data is for $v=0$ with $\epsilon_w/\epsilon =1$,
$r_{cw}=r_c=2^{1/6}\sigma$, $d=1.2\sigma$, $k_B T=1.1\epsilon$,
$N_w=144$, $N_f=288$, and $m_l=2$.
The open and filled circles denote films with $n=1$ and 6, respectively.
Stars indicate the bulk diffusion constant for $n=6$.
Results for $n=20$ (squares) overlap $n=6$ results in (a).
}
\label{fig:loaddep}
\end{figure}
\begin{figure}
\caption{
Effect of varying $P_{\bot}$ on the distribution $\cal D$ of
end-to-end distances, $R_{1n}$, of $n=6$ chains in systems with $v=0$.
All parameters as in Fig. \protect\ref{fig:loaddep}
}
\label{fig:length}
\end{figure}
\begin{figure}
\caption{
Effect of $P_{\bot}$ on chain orientation.
Here $\theta$ is the angle between the $z$-axis and a bond between
adjacent monomers on a chain.
Results for the distribution function $\cal D$
were averaged over 250$\tau$ for the system of
Fig. \protect\ref{fig:loaddep}.
}
\label{fig:theta}
\end{figure}
\begin{figure}
\caption{
Force per unit wall area, $f$, as a function of $v$ at the indicated loads.
Other parameters as in Fig. \protect\ref{fig:loaddep}.
The force saturates at large velocities.
In the limit $v \rightarrow 0$, $f$ goes to 0 when $P_{\bot} < P_{\bot}^G$
and to a constant when $P_{\bot} > P_{\bot}^G$.
}
\label{fig:fvsv}
\end{figure}
\begin{figure}
\caption{
Data of Fig. \protect\ref{fig:fvsv} replotted in (a)
as $\log \mu$ vs. $\log \dot \gamma$.
The corresponding film thickness at each load is shown in (b).
At low $P_{\bot}$ and $\dot \gamma$ the value of $\mu$ is a constant.
Above $\dot \gamma_c$, $\mu$ begins to drop as $\dot \gamma ^{-2/3}$,
and $h$ starts to rise.
For $P_{\bot} > P_{\bot}^G$,
$\mu$ drops as $\gamma ^ {-1}$ at low $\dot \gamma$.
The dashed and dotted lines indicate slopes of $-2/3$ and $-1$, respectively.
}
\label{fig:muvsg}
\end{figure}
\begin{figure}
\caption{
Fit of variation in dynamic quantities $\dot \gamma_c$ and $D$
to free volume theory.
The logarithm of the inverse of both quantities should scale like
$h_0/(h_c-h)$ where $h_c$ is the thickness at $P_{\bot}^G$.
In the fit, $h_0=1.58\sigma$ and $h_c = 1.57 \sigma $,
implying $P_{\bot}^G= 16 \epsilon \sigma^{-3}$.
}
\label{fig:gammac}
\end{figure}
\begin{figure}
\caption{
Data for the system of Fig. \protect\ref{fig:fvsv}, but with
fixed $h$ instead of fixed load.
Note that the data now indicate that $\mu$ falls roughly as
$\dot \gamma ^{-1/2}$ (dashed line) above $\dot \gamma_c$.
}
\label{fig:fixh}
\end{figure}
\begin{figure}
\caption{
Frictional force per unit area $f$ as a function
of $v$ and $n$ at $P_{\bot}=16\epsilon \sigma^{-3}$.
Other parameters as in Fig. \protect\ref{fig:loaddep}.
}
\label{fig:Ndep}
\end{figure}
\begin{figure}
\caption{
Flow profiles for different interaction potentials.
Panel (a) shows results for the system of Fig. \protect\ref{fig:loaddep}
at $v=0.05 \sigma \tau^{-1}$
with $n=1$ (open circles) and $n=6$ (closed circles).
Stars in (b) are for
$n=6$, $m_l=4$, $P_\bot=6 \epsilon \sigma^{-3}$,
$\epsilon_w/\epsilon = 3$, $r_c = 2^{1/6}\sigma$,
$r_{cw} = 2.2\sigma$, and $v=0.02 \sigma\tau^{-1}$
(see also Fig. \protect\ref{fig:muvsg2}).
Squares are for
$n=6$, $m_l=4$, $P_\bot=8 \epsilon \sigma^{-3}$,
$\epsilon_w/\epsilon = 1$, $r_c = 2.2 \sigma$,
$r_{cw} = 2.2\sigma$, and $v=0.08 \sigma \tau^{-1}$
(see also Fig. \protect\ref{fig:muvsg3}).
Particle velocities in the $x$ direction, $v_x$,
were averaged within layers and over at
least $250 \tau$.
These averages were then normalized by the wall velocity $v$
and plotted against the center of each layer $z$ divided by $h$.
The flow profile for a no-slip boundary condition is indicated
by the dotted line in (b).
}
\label{fig:profile}
\end{figure}
\begin{figure}
\caption{
Ratio of the actual shear rate within the film to the value
inferred from a no-slip boundary condition as a function of $n$.
All other parameters as in Fig. \protect\ref{fig:loaddep}.
For $n>3$ only about 15\% of the shear is within the film.
The rest is at the two interfaces.
}
\label{fig:ratio}
\end{figure}
\begin{figure}
\caption{
Plots of $\mu$ vs. $\dot \gamma$ at (a) fixed
$P_{\bot} = 4\epsilon \sigma^{-3}$
and varying $m_l$, and (b) fixed $m_l=4$ and varying $P_{\bot}$.
Dashed lines have slope -2/3.
Both panels show the same behavior as Fig. \protect\ref{fig:muvsg}(a)
as the glass transition is approached.
Other parameters were $\epsilon_w/\epsilon=3$, $r_{cw}=2.2\sigma$,
$k_B T=1.1\epsilon$, $r_c=2^{1/6}\sigma$, $n=6$, and $d=\sigma$.
}
\label{fig:muvsg2}
\end{figure}
\begin{figure}
\caption{
Plot of $\mu$ vs. $\dot \gamma$ at fixed $P_{\bot} = 8\epsilon \sigma^{-3}$
for $m_l=4$, $\epsilon_{w}/\epsilon=1$, $r_{cw}=r_c=2.2\sigma$,
$d=1.2\sigma$, and $k_B T=1.1\epsilon$.
As before, the dashed line has slope -2/3.
}
\label{fig:muvsg3}
\end{figure}
\begin{figure}
\caption{
Time dependence of mean-squared projections on to the $x$ and $y$ axes
of the vectors connecting ends of chains.
The initial state was fully aligned along the $x$-direction.
The ratio of the projections decays to unity by the end of the figure,
showing that the final state is isotropic.
Longer runs and runs from other starting states show the same final isotropy.
}
\label{fig:time}
\end{figure}

\begin{thebibliography}{10}
\begin{references}
\bibitem{AFM1}
C. M. Mate, G. M. McClelland, R. Erlandsson, and S. Chiang,
Phys. Rev. Lett. {\bf 59}, 1942 (1987).
\bibitem{AFM2}
E. Meyer, R. Overney, D. Brodbeck, L. Howald, R. L\"uthi, J. Frommer,
and J.-J. G\"untherodt, Phys. Rev. Lett. {\bf 69}, 1777 (1992).
\bibitem{Krim90}
E. Watts, J. Krim and A. Widom, Phys. Rev. B{\bf 41}, 3466 (1990).
J. Krim, D. H. Solina, R. Chiarello, Phys. Rev. Lett. {\bf 66},  181 (1991).

\bibitem{Jacob88a} J. N. Israelachvili, P. M. McGuiggan, and
    A. M. Homola, Science {\bf 240}, 189 (1988).

\bibitem{McGuiggan89} P. M. McGuiggan, J. N. Israelachvili, M. L. Gee,
    and A. M. Homola,  Mat. Res. Soc. Symp. Proc. {\bf 140}, 79 (1989).

\bibitem{Gee90} M. L. Gee, P. M. McGuiggan, and
    J. N. Israelachvili, J. Chem. Phys. {\bf 93}, 1895 (1990).

\bibitem{VanAlsten88} J. Van Alsten and S. Granick,
    Phys. Rev. Lett. {\bf 61}, 2570 (1988).

\bibitem{VanAlsten90a} J. Van Alsten and S. Granick,
    Langmuir {\bf 6}, 877 (1990).

\bibitem{Hu91} H.-W. Hu, G. A. Carson, and S. Granick,
    Phys. Rev. Lett. {\bf 66}, 2758 (1991).

\bibitem{Granick92} S. Granick, Science {\bf 253}, 1374 (1992).

\bibitem{Jacob90} J. N. Israelachvili, P. M. McGuiggan, M. Gee,
    A. Homola, M. O. Robbins, and P. A. Thompson,
    J. Phys.: Condens. Matter {\bf 2}, SA89 (1990).

\bibitem{Carson92}
G. A. Carson, H.-W. Hu and S. Granick, Tribology Transactions {\bf 35},
405 (1992).

\bibitem{VanAlsten90b} J. Van Alsten and S. Granick,
    Macromolecules {\bf 23}, 4856 (1990).

\bibitem{disks}
A. M. Homola, H. V. Nguyen, and G. Hadziioannou, J. Chem. Phys. {\bf 94},
2346 (1991).
A. M. Homola, G. B. Street, and M. Mate,
{\em MRS Bulletin} {\bf 15} (3), 45 (1990).

\bibitem{Horn81}
R. G. Horn and J. N. Israelachvili, J. Chem. Phys., {\bf 75}, 1400 (1981).

\bibitem{Jacob91} J. N. Israelachvili, Intermolecular and Surface Forces,
2nd ed., (Academic Press, London, 1991).

\bibitem{Abraham78}
F. F. Abraham, J. Chem. Phys. {\bf 68}, 3713 (1978).

\bibitem{Toxvaerd81}
S. Toxvaerd, J. Chem. Phys. {\bf 74}, 1998 (1981).

\bibitem{Plischke86}
M. Plischke and D. Henderson, J. Chem. Phys. {\bf 84}, 2846 (1986).

\bibitem{Bitsanis90b} I. Bitsanis, S. A. Somers, H. T. Davis,
and M. Tirrell, J. Chem. Phys. {\bf 93}, 3427 (1990).

\bibitem{Bitsanis90a} I. Bitsanis and G. Hadziioannou,
J. Chem. Phys. {\bf 92}, 3827 (1990).

\bibitem{Heinbuch89}
U. Heinbuch and J. Fischer, Phys. Rev. A {\bf 40}, 1144 (1989).

\bibitem{Thompson90b} P. A. Thompson and M. O. Robbins,
Phys. Rev. A {\bf 41}, 6830 (1990).

\bibitem{Schoen88}
M. Sch\"oen, J. Cushman, D. Diestler and C. Rhykerd,
J. Chem. Phys. {\bf 88}, 1394 (1988).

\bibitem{Schoen89}
M. Sch\"oen, C. L. Rhykerd, D. Diestler and J. H. Cushman, Science
{\bf 245}, 1223 (1989).

\bibitem{Landman89}
U. Landman, W. D. Luedtke and M. W. Ribarsky,
J. Vac. Sci. Technol. A{\bf 7}, 2829 (1989).

\bibitem{Cieplak94} M. Cieplak, E. D. Smith and M. O. Robbins,
Science {\bf 265}, 1209 (1994).

\bibitem{Thompson90a} P. A. Thompson and M. O. Robbins, Science {\bf 250}, 792
(1990); M. O. Robbins and P. A. Thompson, Science {\bf 253}, 916 (1991).

\bibitem{Thompson92} P. A. Thompson, G. S. Grest and M. O. Robbins,
Phys. Rev. Lett. {\bf 68}, 3448 (1992).

\bibitem{Thompson93a} P. A. Thompson, M. O. Robbins and G. S. Grest,
in {\em Computations for the Nano-Scale}, edited by P.\ Bl\"{o}chl,
C.\ Joachim, and A.J.\ Fisher, (Kluwer, Dordrecht, 1993), p.\ 127.

\bibitem{Thompson93b} P. A. Thompson, M. O. Robbins and G. S. Grest,
in {\em Thin Films in Tribology} edited by D. Dowson, C. M. Taylor and
M. Godet (Elsevier, Amsterdam, 1993), p. 347.

\bibitem{Robbins93} M. O. Robbins, P. A. Thompson and G. S. Grest,
MRS Bulletin {\bf 18} (10), 45-49 (1993).

\bibitem{vanSwol91}
M. Lupowski and F. van Swol, J. Chem. Phys. {\bf 95}, 1995 (1991).

\bibitem{Bitsanis93}
I. Bitsanis and C. Pan, J. Chem. Phys. {\bf 99}, 5520 (1993).

\bibitem{Manias95}
E. Manias, G. Hadziioannou, and G. ten Brinke,
to be published.

\bibitem{Grest81}
G. S. Grest and M. H. Cohen, {\em Advances in Chemical Physics} Vol. 48,
Edited by E. Prigogine and S. A. Rice (Wiley, New York, 1981) p. 455.

\bibitem{Reiter}
G.  Reiter, A. L. Demirel, J Peanasky, L Cai, and S. Granick,
J. Chem. Phys. {\bf 101}, 2606 (1994).
G. Reiter, A. L. Demirel and S. Granick, Science {\bf 263},
1741 (1994).

\bibitem{Kremer90} K. Kremer and G.S. Grest, J. Chem. Phys.
{\bf 92}, 5057 (1990).

\bibitem{Allen87} M. Allen and D. Tildesley,
   {\em Computer Simulation of Liquids} (Clarendon Press, Oxford, 1987).

\bibitem{Grest86} G. S. Grest and K. Kremer, Phys. Rev. A {\bf 33},
3628 (1986).

\bibitem{Horn89} R. G. Horn, S. J. Hirz, G. Hadziioannou,
    C. W. Frank, and J. M. Catala, J. Chem. Phys. {\bf 90},
    6767 (1989).

\bibitem{Magda85}
J. Magda, M. Tirrell and H. T. Davis, J. Chem. Phys. {\bf 83}, 1888 (1985).

\bibitem{Kumar88} A. K. Kumar, M. Vacatello, and D. Y. Yoon,
    J. Chem. Phys. {\bf 89}, 5206 (1988).

\bibitem{Ribarsky92}
M. W. Ribarsky and U. Landman, J. Chem. Phys. {\bf 97}, 1937 (1992).

\bibitem{Xia92}
T. K. Xia, J. Ouyang, M. W. Ribarsky, and U. Landman,
Phys. Rev. Lett. {\bf 69}, 1967 (1992).

\bibitem{Manias93}
E. Manias, G. Hadziioannou, I. Bitsanis, and G. ten Brinke,
Europhys. Lett. {\bf 24}, 99 (1993).

\bibitem{Manias94}
E. Manias, G. Hadziioannou, and G. ten Brinke,
J. Chem. Phys. {\bf 101}, 1721 (1994).

\bibitem{Stevens93}
M. S. Stevens and M. O. Robbins., J. Chem. Phys. {\bf 98}, 2319 (1993).

\bibitem{Lindemann}
See, for example, J. P. Hansen and I. R. McDonald,
Theory of Simple Liquids, 2nd ed., (Academic Press, New York, 1986).

\bibitem{Alsaad78}
M. A. Alsaad, W. O. Winer, F. D. Medina, and D. C. O'Shea,
J. Lubrication Tech. {\bf 100}, 419 (1978).

\bibitem{foot3d}
The results for Fig. \ref{fig:length} are for a number of monomers
that is fixed by $m_l =2$, but at $P_\bot = 1\epsilon \sigma^{-3}$
the film thickness
$h$ is much larger than the thickness of two layers.

\bibitem{Flory69}
P. J. Flory, {\em Statistical Mechanics of Chain Molecules}
(Interscience, New York, 1969).

\bibitem{Carmesin90}
I. Carmesin and K. Kremer, J. Phys. (Paris) {\bf 51}, 915 (1990).

\bibitem{Georges}
J. M. Georges, S. Millot, J. L. Loubet, A. Tonck, and D. Mazuyer,
in {\em Thin Films in Tribology} edited by D. Dowson, C. M. Taylor and
M. Godet (Elsevier, Amsterdam, 1993), p. 443.


\bibitem{Graessley74}
W. W. Graessley, {\em Adv. Polym. Sci.} {\bf 16}, 1 (1974).

\bibitem{Ferry80}
J. D. Ferry, {\em Viscoelastic Properties of Polymers}, 3rd ed.,
(Wiley, New York, 1980).


\bibitem{Bowden58} F.P.Bowden and D. Tabor,
    {\em Friction and Lubrication} (Oxford Univ. Press, Oxford, 1958).

\bibitem{tobe} P. A. Thompson, M. O. Robbins and G. S. Grest,
to be published.

\bibitem{Kroger93} M Kr\"oger, W. Loose and S. Hess,
J. Rheology {\bf 37}, 1057 (1993).

\bibitem{Rabin93}
Y. Rabin and I. Hersht, Physica A{\bf 200}, 708 (1993).

\bibitem{Urbakh95}
M. Urbakh, L. Daikhin and J. Klafter, Phys. Rev. E{\bf 51}, XXXX (1995).

\bibitem{degennes95}
P. G. deGennes, private communication.

\bibitem{Arlet} A. Baljon and M. O. Robbins, to be published.

\end{references}
\end{thebibliography}
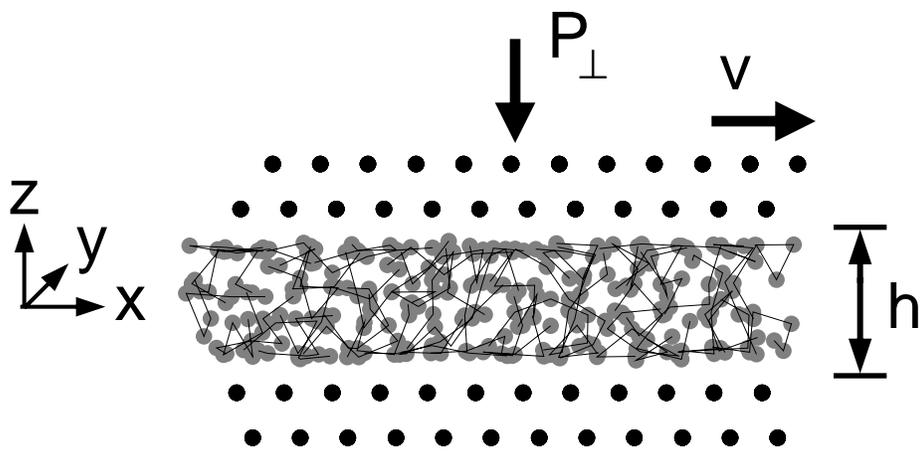
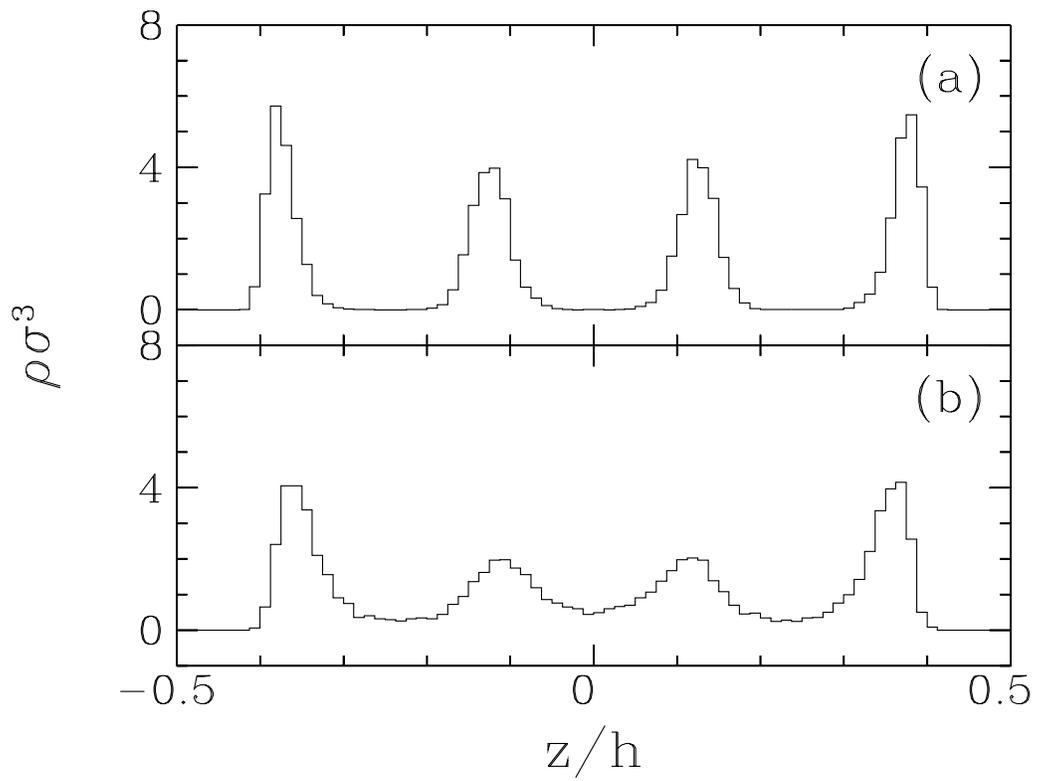
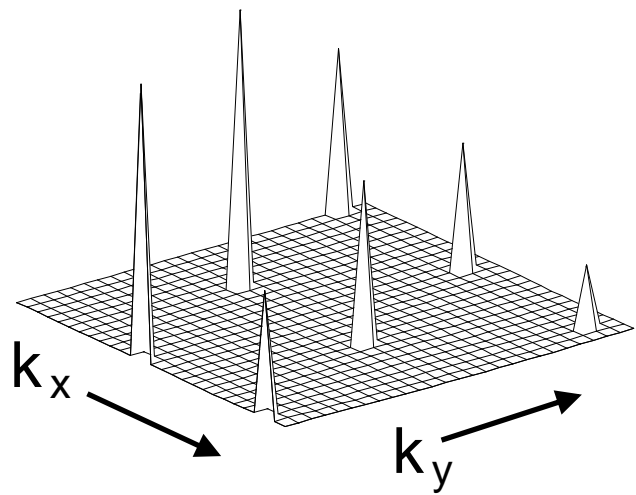
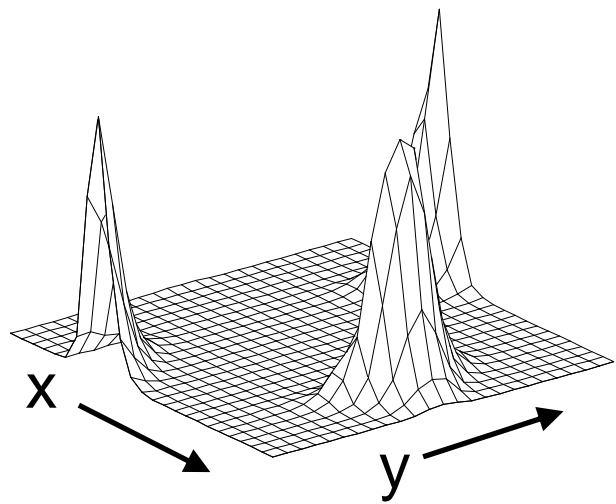
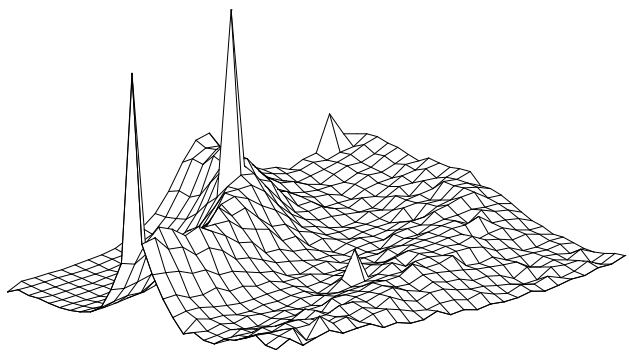
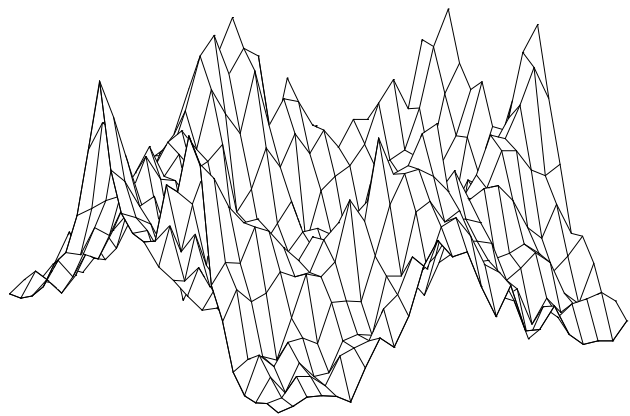
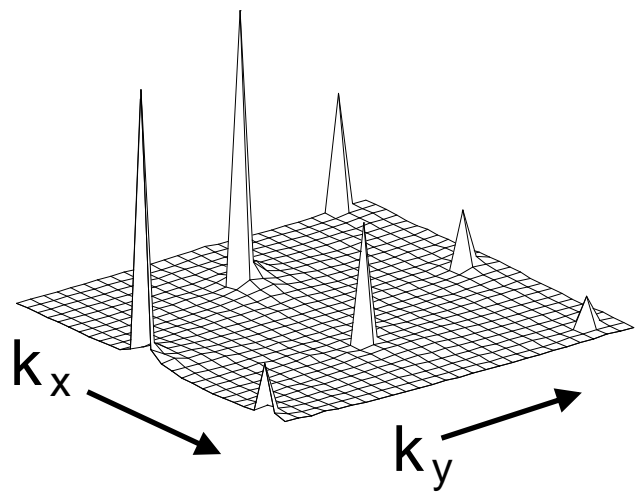
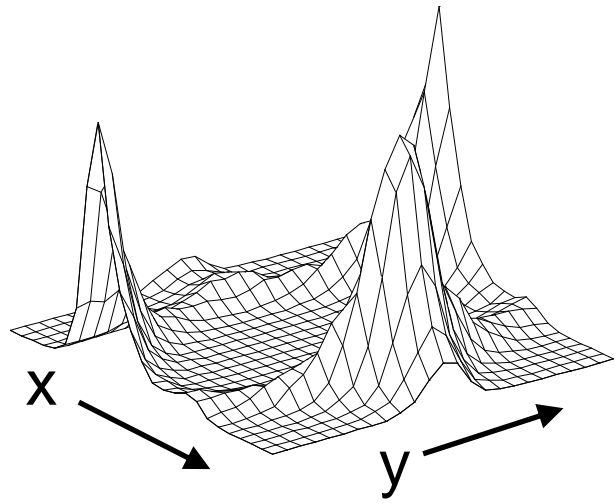
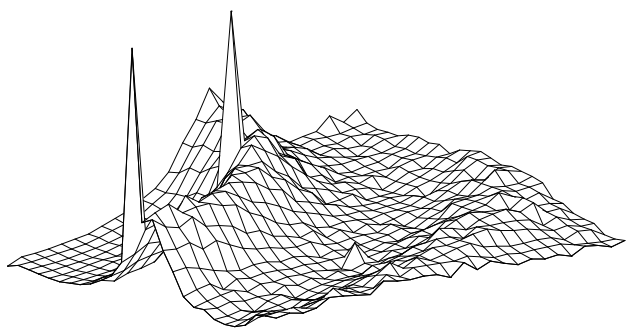
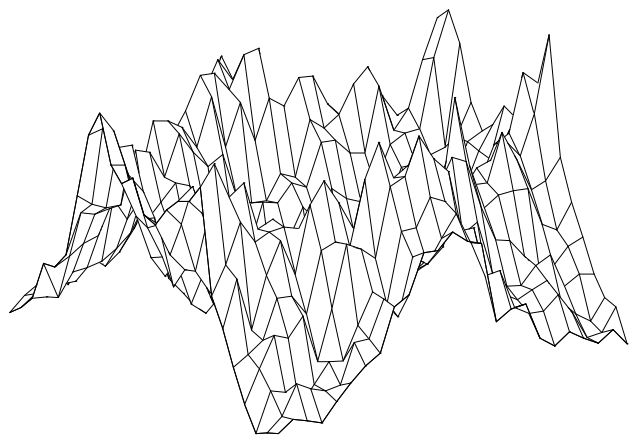
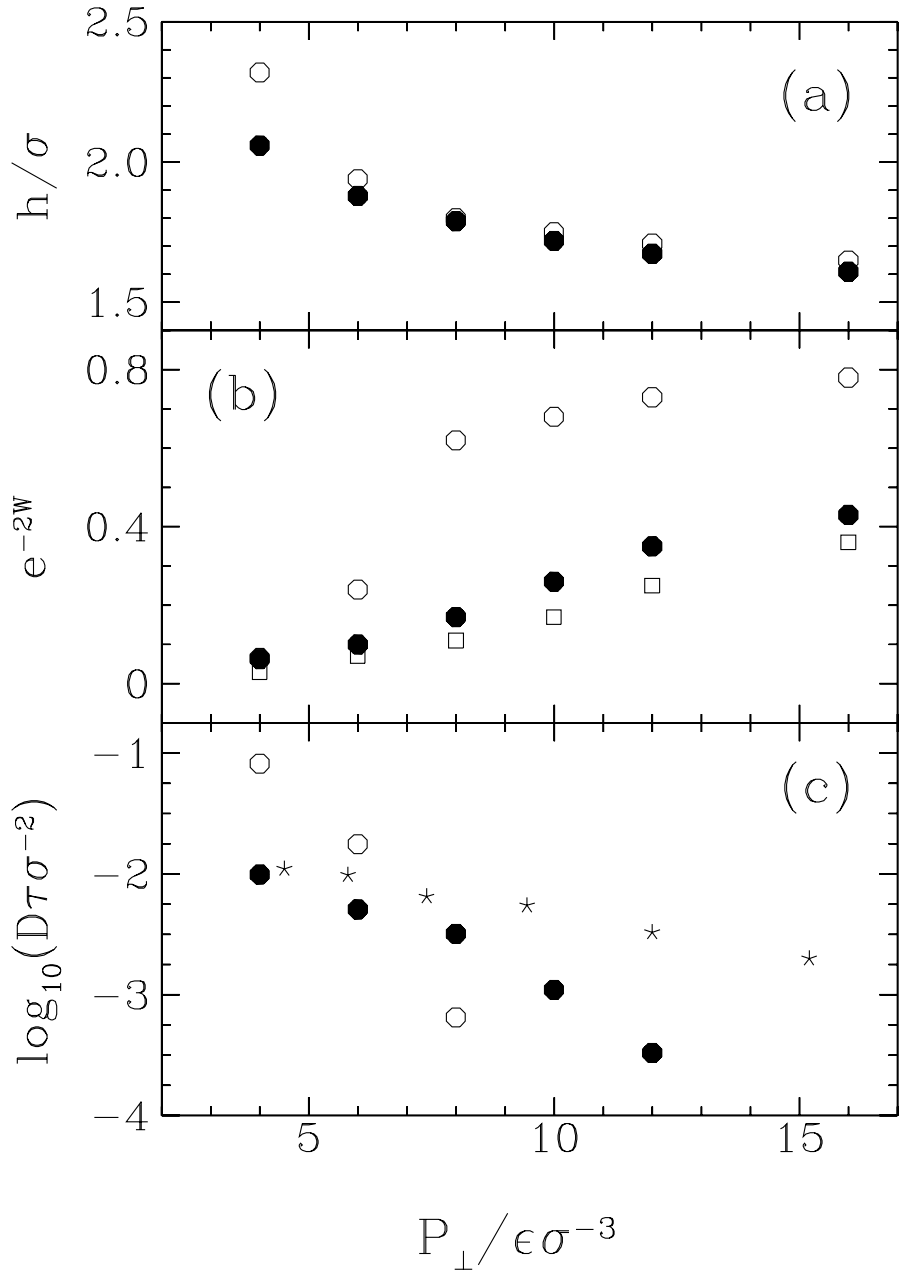
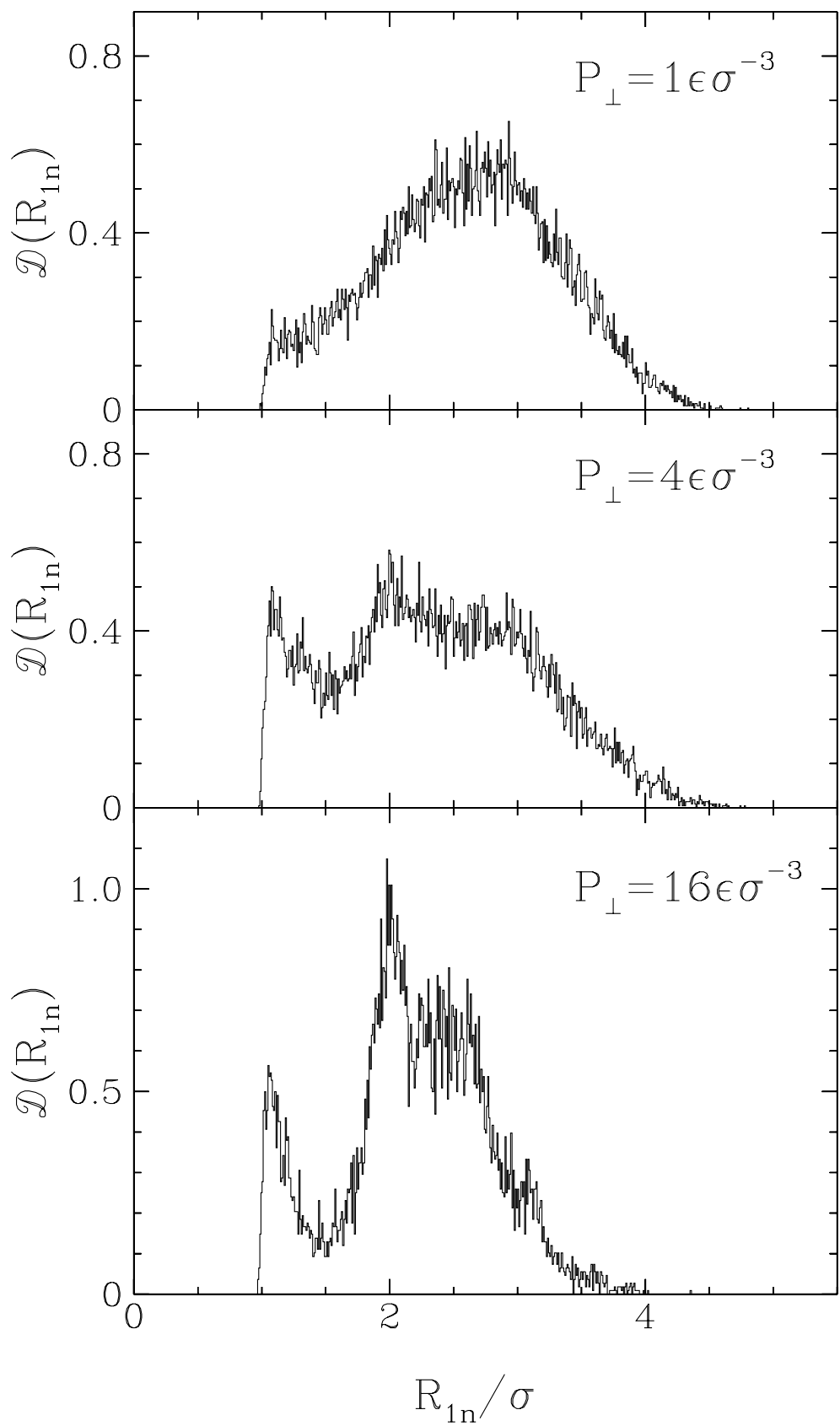
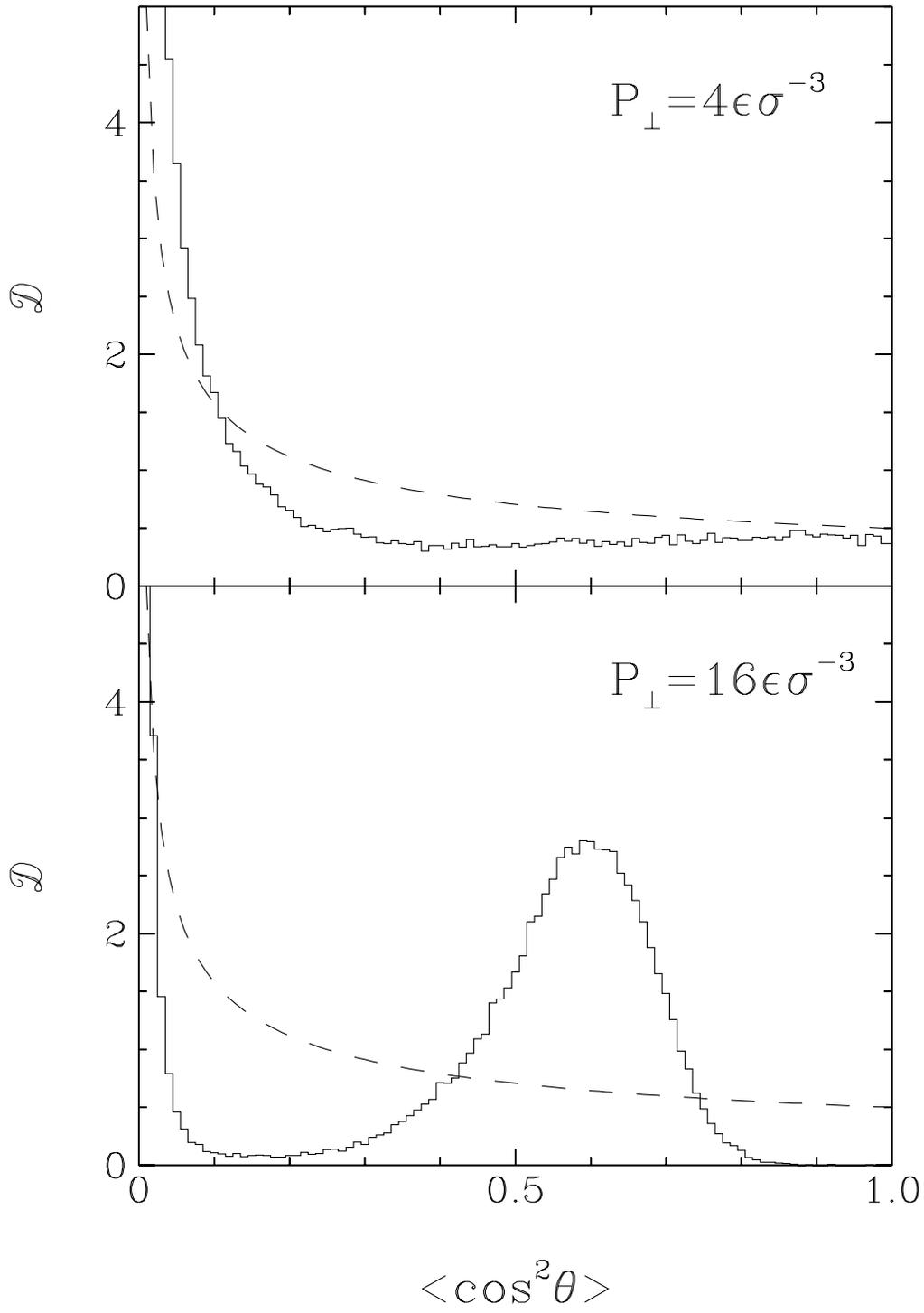
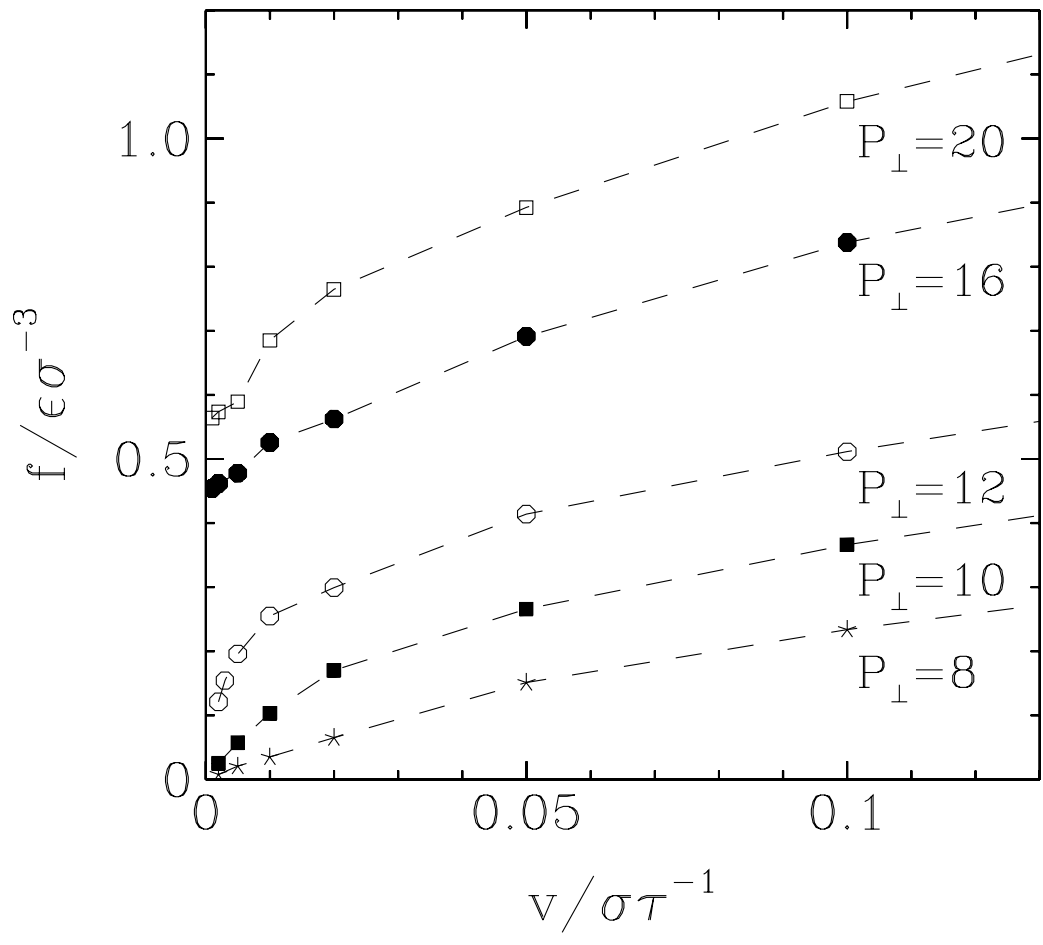
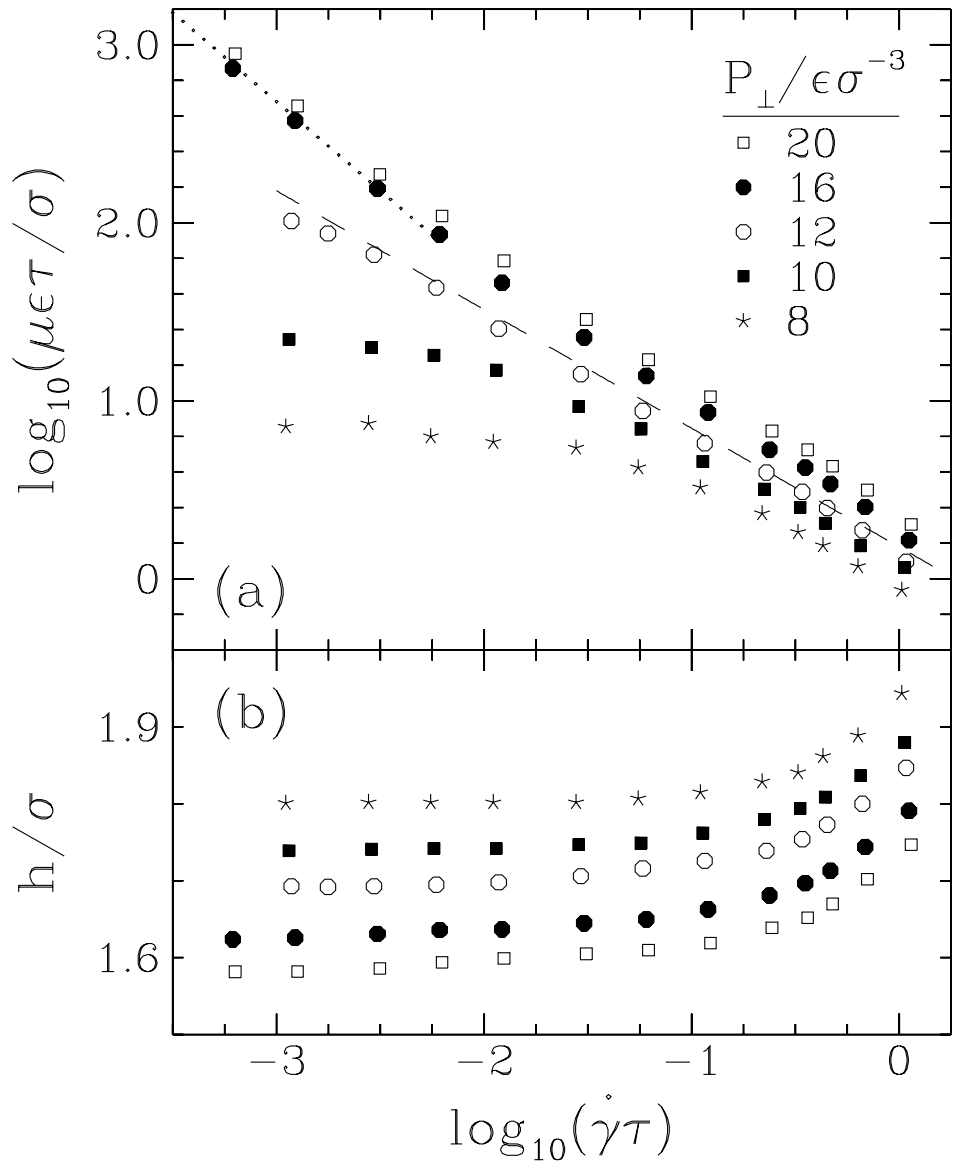
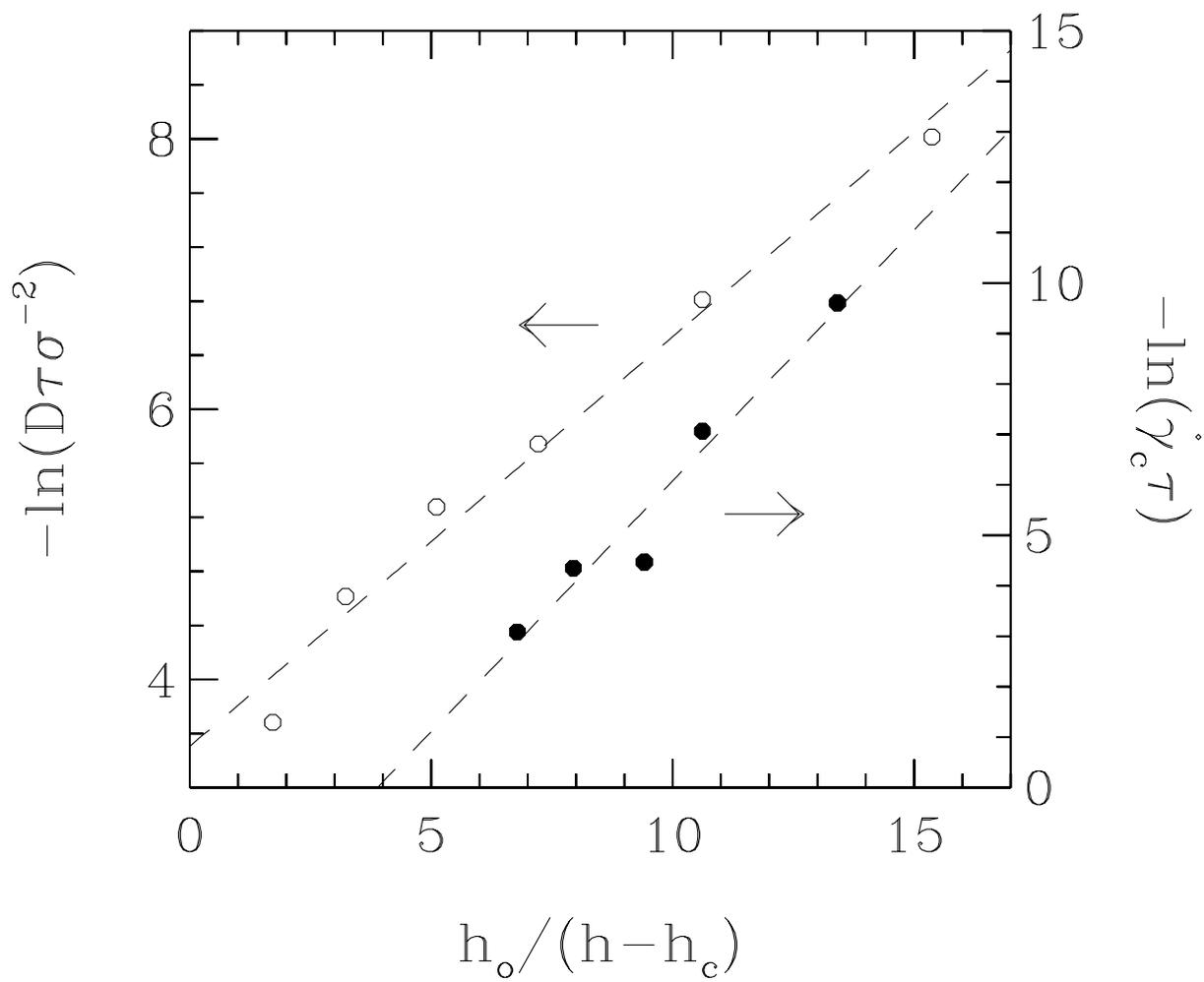
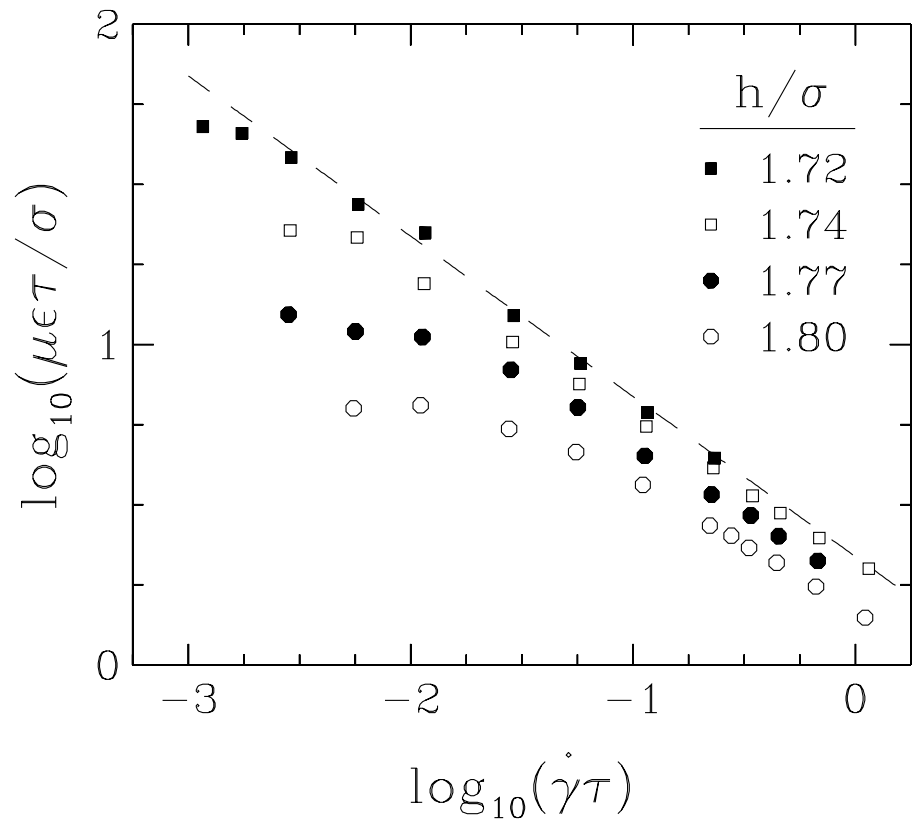
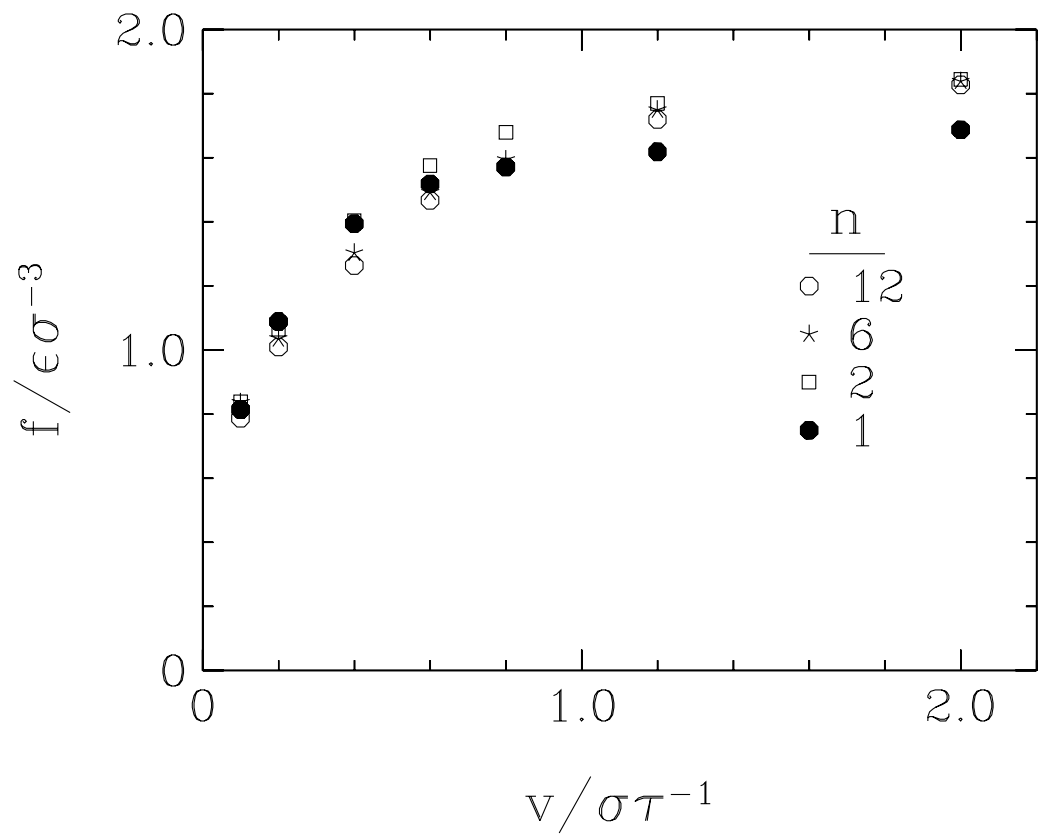
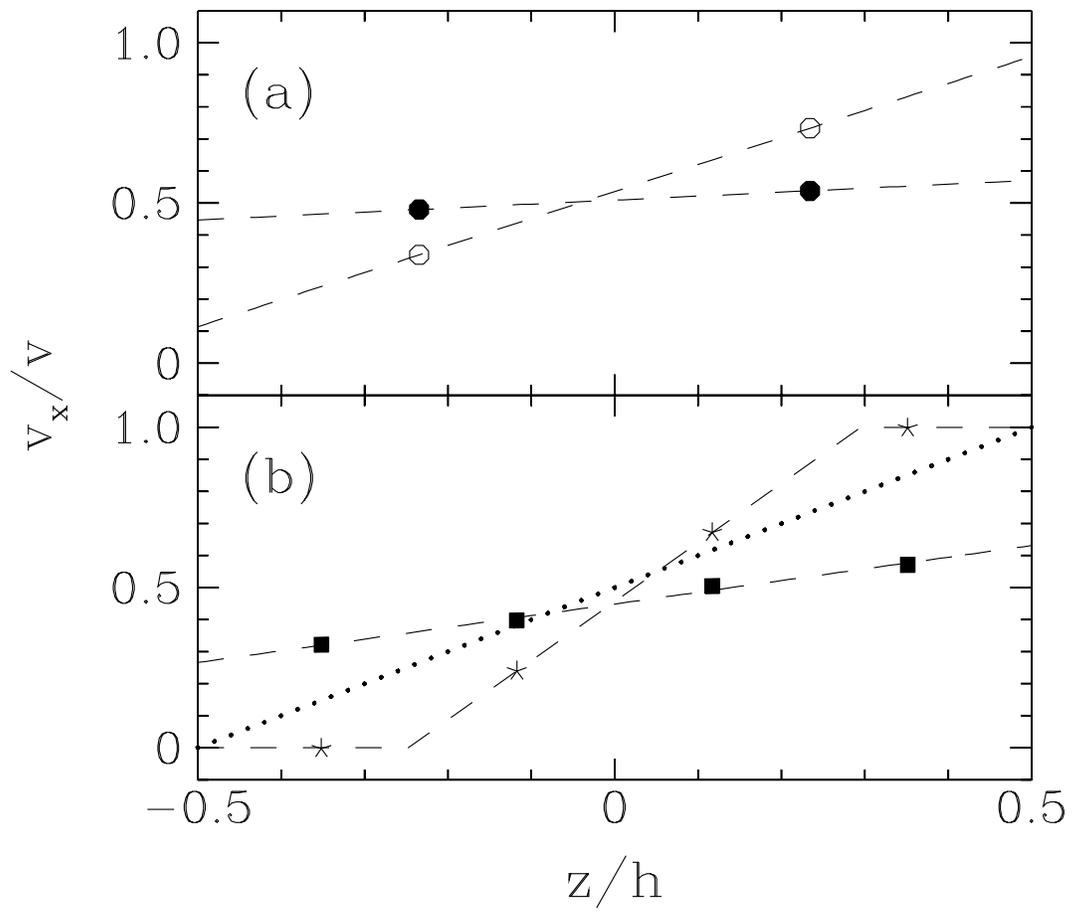
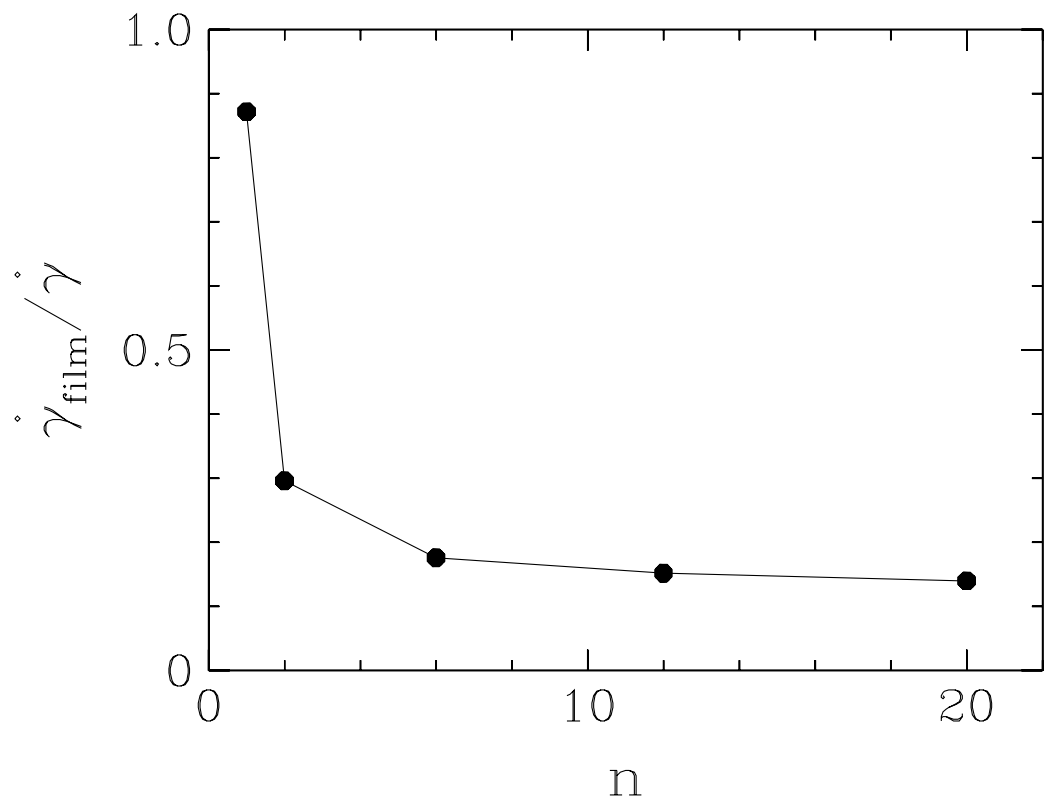
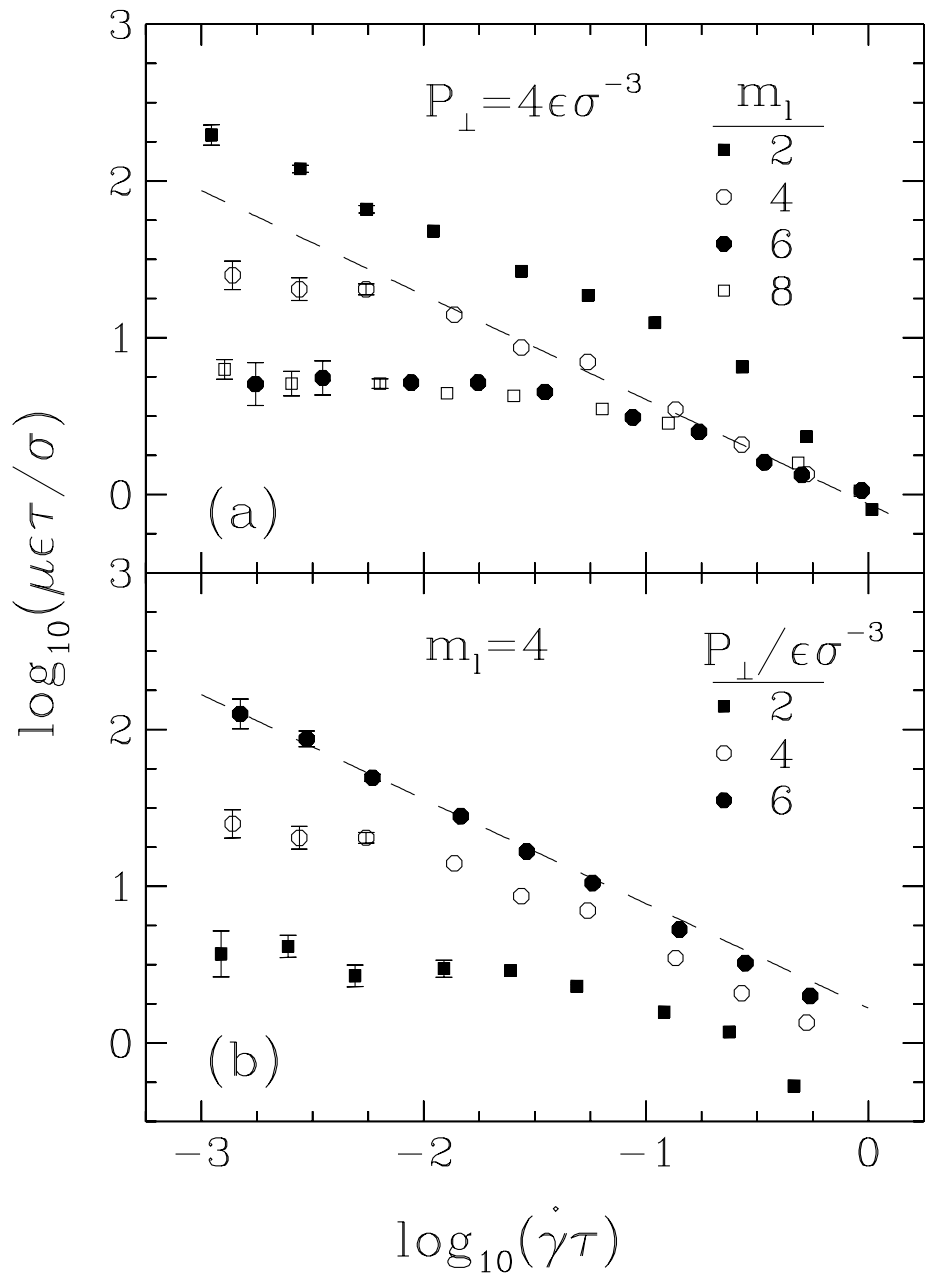
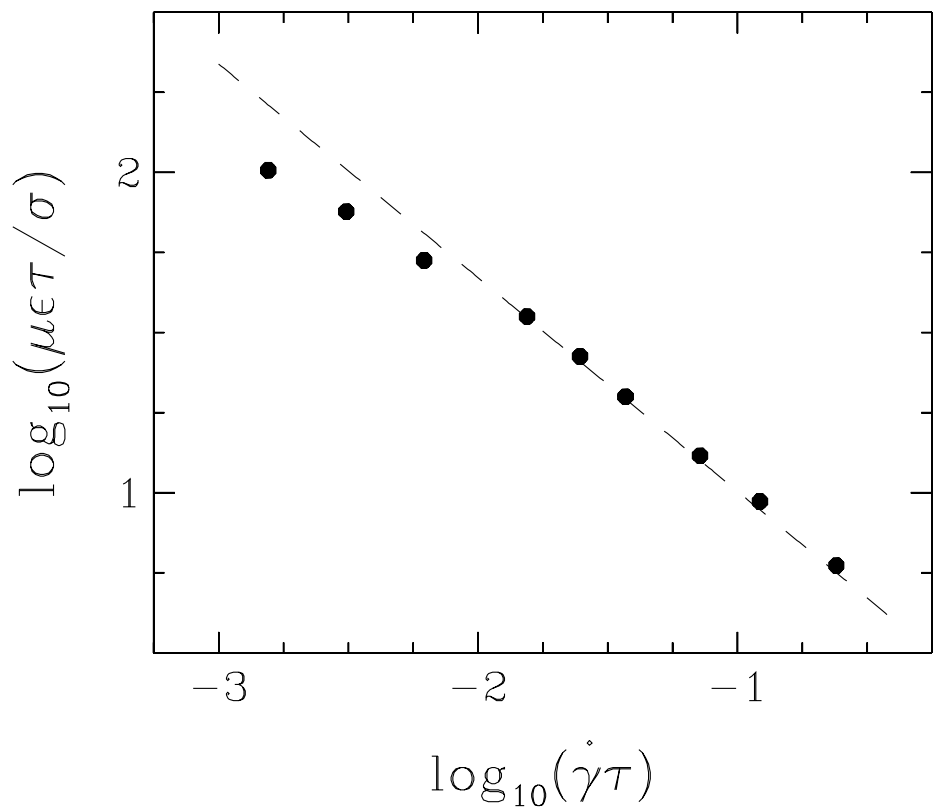
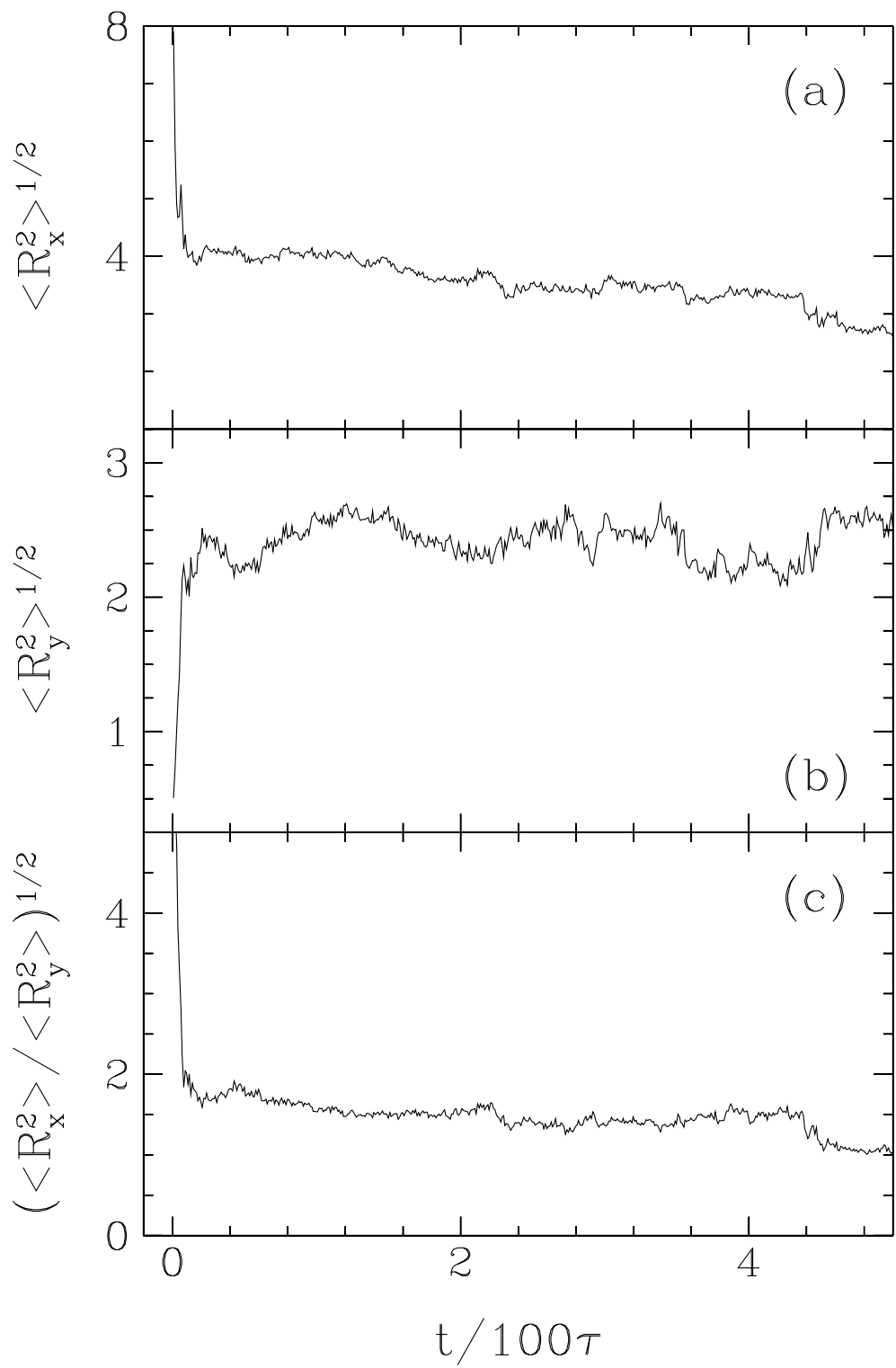
\end{document}